\documentclass[a4paper,fleqn,usenatbib,useAMS]{mnras}
\usepackage{graphicx}	
\usepackage{amsmath}	
\usepackage{amssymb}	
\usepackage{multicol}        
\usepackage{bm}		
\usepackage{pdflscape}	

\usepackage[T1]{fontenc}
\usepackage{ae,aecompl}
\usepackage{txfonts}
\usepackage{threeparttable}

\title[Light curve analysis of RR Lyrae stars]
{On the variation of light curve parameters of RR Lyrae variables at multiple wavelengths}
\author[Das et al.]{Susmita Das$^{1}$\thanks{E-mail:susmitadas130@gmail.com}, Anupam Bhardwaj$^{2}$, Shashi M. Kanbur$^{3}$, Harinder P. Singh$^{1}$  and  \and Marcella Marconi$^{4}$\\
\vspace{1pt}\\
1. Department of Physics \& Astrophysics, University of Delhi, Delhi 110007, India \\
2. Kavli Institute for Astronomy and Astrophysics, Peking University, Yi He Yuan Lu 5, Hai Dian District, Beijing 100871, China\\
3. State University of New York, Oswego, NY, USA \\
4. INAF-Osservatorio astronomico di Capodimonte, Via Moiariello 16, 80131 Napoli, Italy
}

\begin{document}

\date{Accepted 2018 August 24. Received 2018 August 24; in original form 2018 May 16}

\pagerange{\pageref{firstpage}--\pageref{lastpage}} \pubyear{2018}

\maketitle

\label{firstpage}

\begin{abstract}
We present a detailed light curve analysis of RR Lyrae variables at multiple wavelengths using Fourier decomposition method. The time-series data for RR Lyrae variables in the Galactic bulge and the Magellanic Clouds are taken from the Optical Gravitational Lensing Experiment survey while the infrared light curves are compiled from the literature. We also analyse the multiband theoretical light curves that are generated from the stellar pulsation models of RR Lyrae stars for a wide range of metal-abundances. We find that the theoretical light curve parameters with different metal abundances are consistent with observed parameters in most period bins at both optical and infrared wavelengths. The theoretical and observed Fourier amplitude parameters decrease with increase in wavelength while the Fourier phase parameters increase with wavelength at a given period. We use absolute magnitudes for a subset of theoretical models that fit the observed optical RR Lyrae light curves in the Large Magellanic Cloud to estimate a distance modulus, $\mu_\textrm{LMC}=18.51\pm0.07$, independent of the metallicity. We also use Fourier analysis to study the period-color and amplitude-color relations for RR Lyrae stars in the Magellanic Clouds using optical data and find that the slope of period-color relation at minimum light is very shallow or flat and becomes increasingly significant at the maximum light for RRab stars. We also find that the metallicity dependence of the period-color relations increases as we go from minimum to maximum light, suggesting that the mean light results are indeed an average of the various pulsational phases. We summarize that the average variation in these relations is consistent between theory and observations and supports the theory of the interaction of the stellar photosphere and the hydrogen ionization front.
\end{abstract}

\begin{keywords}
stars: variables: RR Lyrae- Galaxy: bulge - galaxies: Magellanic Clouds
\end{keywords}

\section{Introduction}

RR Lyrae stars are low-mass, helium-burning, horizontal branch stars, exhibiting periodic light curves with a pulsation period of $\sim0.2-1.0$~day and amplitude variation of $\le$~2 mag \citep{preston1964, kolenberg2012}. These radially pulsating variables are valuable tracers of old and metal-poor stellar populations and provide extragalactic distance estimates with well-defined period-luminosity relations ($PLR$s), especially in the near-infrared bands \citep{longmore1986, bono2001, catelan2004, sollima2006, muraveva2015, braga2015, neeley2015}. Additionally, a narrow range of the intrinsic colors of RR Lyrae stars also serves as a useful reddening indicator \citep{sturch1966, mateo1995, guldenschuh2005,wagner-kaiser2017}. RR Lyrae and Cepheid variables are both excellent probes for the understanding of the theory of stellar pulsation and evolution. A comparison of their observed light curves and pulsation properties with theory provides very useful constraints for the stellar pulsation models \citep{simon1985,marconi2013,marconi2017,bhardwaj2017}.

Light curve structures of Cepheid and RR Lyrae variables were first studied using the Fourier decomposition method by \citet{simon1981} and \citet{simon1982}, respectively. The variation of lower-order Fourier parameters with period for 70 field RR Lyrae stars was discussed by \citet{simon1982}. They found that fundamental-mode (RRab) and first-overtone (RRc) mode RR Lyrae stars can be easily differentiated on the Fourier amplitude plane. \citet{petersen1984} carried out a Fourier analysis of RR Lyrae stars in the $\omega$ Centauri and found evidence of Cepheid-like progressions \citep{simon1981} in RRab stars in the period range 0.5 day to 1.5 days. \citet{simon1985} compared the light curves of RR Lyrae stars with those from hydrodynamical models using Fourier decomposition and while they found consistent results between theoretical and observed data for the RRc stars, there was a discrepancy in the Fourier phase parameters for RRab stars. \citet{bono1996} computed pulsation models of RR Lyrae stars with different chemical compositions and found that the bolometric amplitudes increase for RRab stars and decrease for RRc stars with an increase in the metal abundance. Using a smooth grid of models covering a range of stellar masses, luminosities and metallicities, \citet{bono2001} provided theoretical constraints on the $PLR$ for RR Lyrae stars in the $K$-band. Most of these theoretical studies focussed on the comparison with observed pulsation properties \citep{marconi2011,bono2011,bono2016,marconi2016}, while others carried out a model fitting of the light curves e.g. \citet{simon1985,wood1997,bono2002,natale2008,marconi2010,marconi2013,marconi2013b,marconi2017} for Cepheids and \citet{kovacs1998,bono2000,castellani2002,difabrizio2002,marconi2005,marconi2007} for RR Lyrae stars. However, this work makes use of the modern time-series data and the most recent stellar pulsation models of RR Lyrae stars from Marconi et al. (2015) to provide an extensive comparison of theoretical and observed light curves of RR Lyrae variables.

Fourier analysis of RR Lyrae stars has also been employed to obtain photometric metallicities using the empirical light curve structure and metallicity relations \citep{kovacs1995}. \citet{jurcsik1996} investigated the best-fit relations between $[Fe/H]$ and the Fourier parameters and found a linear relation among $[Fe/H]$, period and the Fourier phase parameter ($\phi_{31}$). \citet{morgan2007} found a $[Fe/H]-\phi_{31}-P$ relation for RRc stars in the globular clusters analogous to that found by \citet{jurcsik1996} for RRab stars. The correlation of Fourier phase parameters with metallicity for RR Lyrae stars has been revisited by several authors, for example, \citet{smolec2005, nemec2013, skowron2016, ngeow2017} and references therein. These empirical relations allow studies of morphology, structure and the metallicity of the Galaxy and the Magellanic Clouds \citep{haschke2012,deb2014,deb2015,hajdu2015,skowron2016}.

Additionally, light-curve data of RR Lyrae stars can also be used to study period-color (PC) and amplitude-color (AC) relations as a function of pulsation phase to understand the interaction of the stellar photosphere and the hydrogen ionization front \citep{simon1993, kanbur2005, bhardwaj2014, ngeow2017}. Recent near-infrared observations of RR Lyrae stars provide evidence of a tight $PLR$ at longer wavelengths \citep{sollima2006, principe2006, coppola2012, muraveva2015, braga2015, neeley2017} that can be used to measure the value of the Hubble constant up to a few percent precision \citep{beaton2016}. However, metallicity contribution to $PLR$s for RR Lyrae stars is not well-constrained as the observed $P-L-[Fe/H]$ relations differ from those obtained using theoretical pulsation models \citep[for example,][]{muraveva2015,marconi2015}. A detailed light curve analysis of RR Lyrae stars can provide insights into the metallicity effects on the light curve structure and subsequently on the mean light PLRs. Further, a comparison of the observed light curves of RR Lyrae stars with theoretical models is crucial for understanding important constraints for the stellar pulsation codes. This analysis has been recently carried out for Cepheid variables by \citet{bhardwaj2017} and we extend this work for RR Lyrae stars in the present analysis.

\begin{table}
\caption{A summary of 410 RR Lyrae models (274 RRab and 136 RRc) used in the present analysis. The last two columns denote the number of RRab/RRc models with a unique combination of (Z,Y,$\dfrac{M}{M_{\odot}}$,$\log\dfrac{L}{L_{\odot}}$). ${T_e}$ indicates the range of effective temperature for each of these combinations.}
\label{tab:Data}
\centering
\begin{tabular}{c c c c c c c}
\hline
\hline
Z & Y & $\dfrac{M}{M_{\odot}}$ & $\log\dfrac{L}{L_{\odot}}$ & ${T_e}$ (K) & RRab & RRc
\\ [0.5ex]
\hline
\hline
& & 0.51 & 1.69 & 5700$-$6800 & 7 & 3 \\[-1ex]
& & 0.51 & 1.78 & 5600$-$6800 & 7 & 3 \\[-1ex]
\raisebox{1ex}{0.02} & \raisebox{1ex}{0.27} & 0.54 & 1.49 & 6000$-$7100 & 5 & 3\\[-1ex]
& & 0.54 & 1.94 & 5200$-$6600 & 8 & $-$ \\[1ex]

& & 0.55 & 1.62 & 5900$-$7100 & 12 & 4 \\[-1ex]
& & 0.55 & 1.72 & 5800$-$7000 & 12 & 10 \\[-1ex]
& & 0.56 & 1.60 & 5900$-$7100 & 11 & 4 \\[-1ex]
\raisebox{1ex}{0.008} & \raisebox{1ex}{0.256} & 0.56 & 1.70 & 5800$-$7000 & 10 & 9 \\[-1ex]
& & 0.57 & 1.58 & 6000$-$7100 & 5 & 3\\[-1ex]
& & 0.57 & 2.02 & 5400$-$6680 & 6 & $-$\\[1ex]

& & 0.53 & 1.81 & 5700$-$6800 & 8 & 4 \\[-1ex]
& & 0.55 & 1.71 & 6000$-$7000 & 10 & 8 \\[-1ex]
& & 0.55 & 1.81 & 5700$-$6900 & 13 & $-$ \\[-1ex]
& & 0.56 & 1.65 & 6000$-$7100 & 10 & 4 \\[-1ex]
\raisebox{0ex}{0.004} & \raisebox{0ex}{0.25} & 0.56 & 1.75 & 5800$-$7000 & 10 & 8 \\[-1ex]
& & 0.57 & 1.63 & 6000$-$7100 & 10 & 4\\[-1ex]
& & 0.57 & 1.73 & 5900$-$7000 & 10 & 8\\[-1ex]
& & 0.59 & 1.61 & 6000$-$7200 & 10 & 3\\[-1ex]
& & 0.59 & 2.02 & 5700$-$6700 & 7 & $-$\\[1ex]

& & 0.58 & 1.87 & 5900$-$6900 & 7 & 4 \\[-1ex]
\raisebox{0ex}{0.001} & \raisebox{0ex}{0.245} & 0.64 & 1.67 & 6000$-$7200 & 9 & 4 \\[-1ex]
& & 0.64 & 1.99 & 5700$-$6800 & 10 & 5\\[1ex]

& & 0.6 & 1.89 & 5700$-$6900 & 9 & 7 \\[-1ex]
\raisebox{0ex}{0.0006} & \raisebox{0ex}{0.245} & 0.67 & 1.69 & 6000$-$7200 & 9 & 6 \\[-1ex]
& & 0.67 & 2.01 & 5800$-$6800 & 9 & 6\\[1ex]

& & 0.65 & 1.92 & 5800$-$6900 & 6 & 7 \\[-1ex]
\raisebox{0ex}{0.0003} & \raisebox{0ex}{0.245} & 0.716 & 1.72 & 6000$-$7200 & 8 & 7 \\[-1ex]
& & 0.716 & 1.99 & 5700$-$6900 & 11 & 3\\[1ex]

& & 0.72 & 1.96 & 5800$-$6900 & 7 & 2 \\[-1ex]
\raisebox{0ex}{0.0001} &  \raisebox{0ex}{0.245} & 0.8 & 1.76 & 6000$-$7200 & 8 & 7\\[-1ex]
& & 0.8 & 1.97 & 5800$-$6700 & 10 & $-$\\[1ex]
\hline
\end{tabular}
\end{table}

The structure of this paper is as follows: Section \ref{sec:data} describes the theoretical models and the observational data used in this analysis. The Fourier decomposition technique is briefly discussed in Section \ref{sec:fourier} and we study the variation of amplitude and Fourier parameters with wavelength, period and metallicity and present the comparison of theoretical and observed light curve parameters. We discuss the extinction corrected PC, AC and PC-metallicity relations for RR Lyrae stars in Section \ref{sec:PCAC}. Finally, we summarise the results of this study in Section \ref{sec:results}.

\begin{table*}
\caption{The observed light curve data used in the present analysis with the number of stars available in each dataset. N$_\textrm{RRab}$, N$_\textrm{RRc}$ and N$_\textrm{T2C}$ refer to the number of RRab, RRc and type II Cepheid variables.}
\centering
\begin{tabular}{c c c c c c c}
\hline\hline
& Band & N$_\textrm{RRab}$  & N$_\textrm{RRc}$	& Reference  & N$_\textrm{T2C}$	& Reference\\
\hline \hline
& V & 11736  & 5603 & & 517 & \\[-1ex]
\raisebox{1ex}{Bulge} & & & & \raisebox{1ex}{\citet{soszynski2014}} & & \raisebox{1ex}{\citet{soszynski2017}}\\[-1ex]
& I & 25929  & 10261 &  & 873 & \\[1ex]
& V & 25762  & 8758 & & 202 & \\[-1ex]
\raisebox{1ex}{LMC} & & & & \raisebox{1ex}{\citet{soszynski2016}} & & \raisebox{1ex}{\citet{soszynski2008}}\\[-1ex]
& I & 26209  & 8893 &  & 203 & \\[1ex]
& V & 4769  & 739 & & 42 & \\[-1ex]
\raisebox{1ex}{SMC} & & & & \raisebox{1ex}{\citet{soszynski2016}} & & \raisebox{1ex}{\citet{soszynski2010}}\\[-1ex]
& I & 4825  & 746 &  & 43 & \\[1ex]
& 3.6 $\mu$m & 14 &  $-$ & & $-$ & \\[-1ex]
\raisebox{1ex}{M4} & & & & \raisebox{1ex}{\citet{neeley2015}} & & \raisebox{1ex}{$-$}\\[-1ex]
& 4.5 $\mu$m & 14 &  $-$ &  & $-$ & \\[1ex]
\hline
\end{tabular}
\label{tab:Observed_Data}
\end{table*}

\begin{table*}
\caption{The light curve parameters of RR Lyrae stars from the theoretical models. The columns provide the chemical composition (Z and Y), filter ($\lambda$), stellar mass, mode of pulsation, effective temperature ($T_e$), logarithmic luminosity, logarithmic period, amplitude ($A$), mean magnitude ($m_0$), Fourier amplitude ($R_{21}$,$R_{31}$) and phase ($\phi_{21}$,$\phi_{31}$) parameters and the mean radius.}
\centering
\scalebox{0.95}{
\begin{tabular}{c c c c c c c c c c c c c c c}
\hline\hline
Z & Y & $\lambda$ & $\frac{M}{M_{\odot}}$ & Mode & $T_e$ & $\log\frac{L}{L_{\odot}}$ & $\log(P)$ & A & $m_0$ & $R_{21}$ & $R_{31}$ & $\phi_{21}$ & $\phi_{31}$ & $\log\frac{R}{R_{\odot}}$\\
[0.5ex]
\hline \hline
0.0200&	0.270&	U&	0.51&	FU&	6800&	1.69&	-0.193&	1.258&	1.056&	0.407&	0.152&	2.838&	5.627&	0.702\\
0.0200&	0.270&	U&	0.51&	FU&	6700&	1.69&	-0.171&	1.661&	1.123&	0.494&	0.176&	3.041&	6.006&	0.715\\
0.0200&	0.270&	U&	0.51&	FU&	6500&	1.69&	-0.127&	1.596&	1.192&	0.474&	0.089&	3.463&	0.594&	0.741\\
0.0200&	0.270&	U&	0.51&	FU&	6200&	1.69&	-0.058&	1.044&	1.293&	0.458&	0.131&	4.462&	1.993&	0.782\\
0.0200&	0.270&	U&	0.51&	FU&	6500&	1.69&	-0.127&	1.596&	1.192&	0.474&	0.089&	3.463&	0.594&	0.741\\
0.0200&	0.270&	U&	0.51&	FU&	5900&	1.69&	0.012&	0.433&	1.431&	0.871&	0.205&	4.722&	1.338&	0.826\\
0.0200&	0.270&	U&	0.51&	FU&	5700&	1.69&	0.064&	0.519&	1.592&	0.448&	0.224&	5.107&	1.510&	0.856\\
0.0200&	0.270&	U&	0.54&	FU&	6800&	1.49&	-0.379&	1.703&	1.586&	0.476&	0.283&	2.807&	5.510&	0.603\\
0.0200&	0.270&	U&	0.54&	FU&	6700&	1.49&	-0.356&	1.481&	1.604&	0.449&	0.243&	2.848&	5.662&	0.615\\
0.0200&	0.270&	U&	0.54&	FU&	6500&	1.49&	-0.313&	1.169&	1.653&	0.388&	0.196&	3.028&	6.215&	0.641\\
\hline
\end{tabular}}
\begin{tablenotes}
      \small
      \item \textbf{Notes:} This table is available entirely in a machine-readable form in the online journal as supporting information.
    \end{tablenotes}
\label{tab:Models}
\end{table*}

\section{The Data}
\label{sec:data}
\subsection{Theoretical light curve data}
We analyse the theoretical light curves generated using the nonlinear, time-dependent convective hydrodynamical models of 243 RR Lyrae stars (166 RRab, 77 RRc) from \citet{marconi2015}, computed with a constant helium-to-metal abundance ratio. 167 additional models (108 RRab, 59 RRc) with different metal abundances, stellar masses and luminosities have been computed for the present analysis and the summary of the models available is listed in Table~\ref{tab:Data}. The models have seven different chemical compositions ranging from Z=0.02 to Z=0.0001, with a primordial He abundance of 0.245 and a helium-to-metals enrichment ratio of 1.4. The $[He/M]$ value has been adopted to correctly replicate the initial helium abundance of the Sun \citep{serenelli2010}. We note here that $\textrm{Z}_{\odot}$=0.0122 \citep{asplund2005}. Each of these compositions have a few sets of stellar masses and luminosities, which were fixed according to detailed central He-burning horizontal-branch evolutionary models. The typical range of Z is broad enough for the comparison with the observed RR Lyrae stars in
the Galaxy, LMC and SMC \citep{clementini2003}. For a fixed Z, the range of M,L,T is certainly not smooth enough to cover all possible combinations. Regardless, these set of models will be used to provide an overall consistency test for the comparison with observed light curves structure. However, we emphasize that a smoother grid of models may allow us to study the impact of different physical parameters on the light curve structure. The predicted bolometric light curves for both fundamental (FU) and first overtone pulsators (FO) have been transformed into optical ($UBVRI$) and near-infrared (NIR;$JKL$) bands using static model atmospheres \citep{bono1995,castelli1997,castelli1997b}. The models also include RR Lyrae stars with periods greater than 1 day, taking into account the possibility of evolved RR Lyrae stars.

\subsection{Archival observed light curve data}
We also use archival observed light curve data for a comparison with the theoretical results. This is summarised in Table~\ref{tab:Observed_Data}. The optical ($VI$) light curves are taken from the OGLE-IV catalog of RR Lyrae variables in the Large Magellanic Cloud (LMC) and Small Magellanic Cloud (SMC) \citep{soszynski2016} and the Galactic Bulge \citep{soszynski2014}. The mid-infrared (3.6 $\mu$m and 4.5 $\mu$m) time-series photometry data for RR Lyrae stars in the globular cluster M4 (NGC 6121) is taken from \citet{neeley2015}. We have also included the data of type II Cepheids in the LMC \citep{soszynski2008}, SMC \citep{soszynski2010} from the OGLE-III catalog and the Galactic Bulge \citep{soszynski2017} from the OGLE-IV catalog in the optical ($VI$) bands for a comparison with the models of evolved RR Lyrae stars with different chemical compositions. 

\begin{figure}
\centering
\includegraphics[scale = 1,width=1.0\columnwidth]{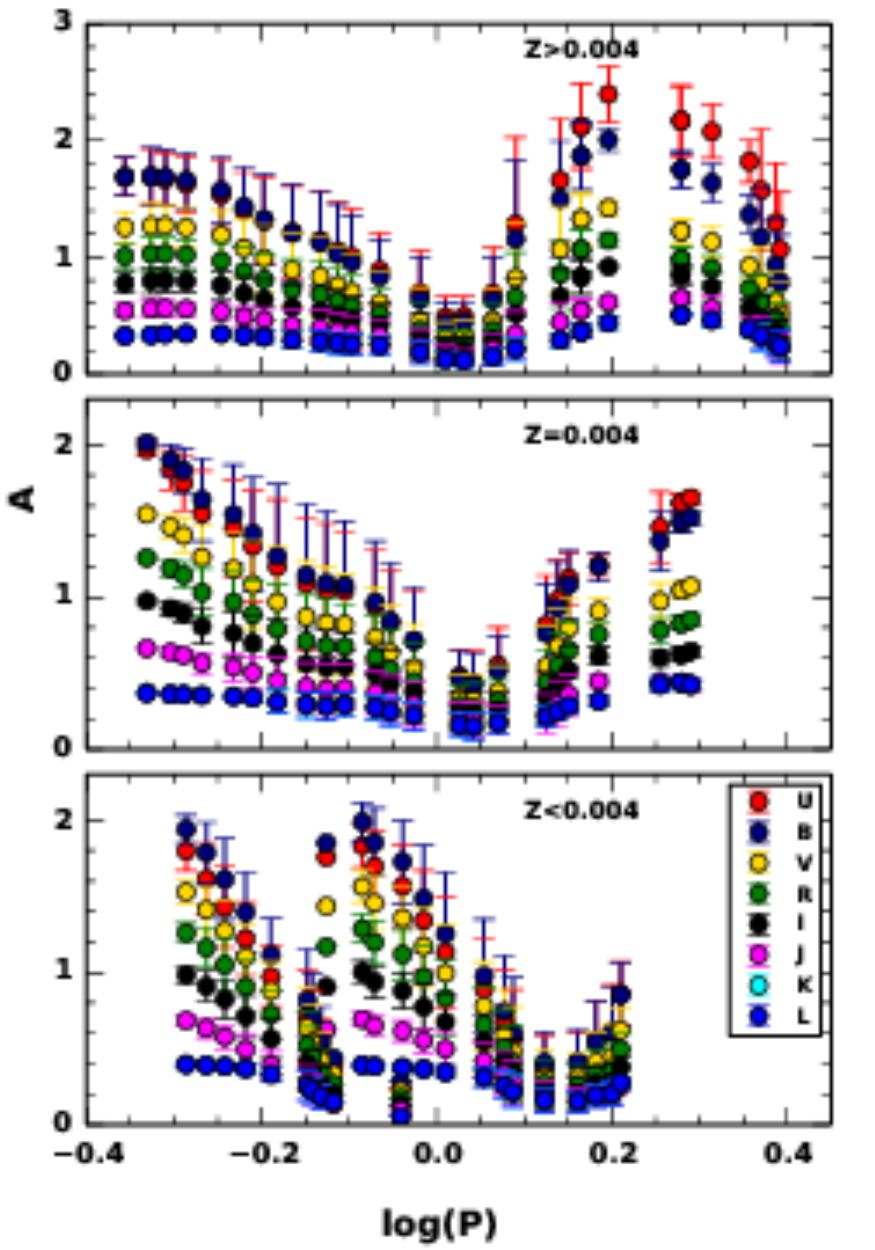}
\caption{The variation of the mean amplitudes (A) as a function of period and wavelength for the theoretical light curves of RRab stars with metal abundance Z$>$0.004, Z=0.004 and Z$<$0.004. Error bars are standard deviations on the mean value within a given bin size.}
\label{fig:A-P}
\end{figure}

In order to study the pulsation properties of the observed RR Lyrae stars at a particular phase or mean-light, we need to account for the extinction corrections in magnitudes at a given wavelength. Using the positions (RA/Dec), we obtain the color excess $E(V-I)$ values for RR Lyrae stars and type II Cepheids in the LMC and SMC from the reddening maps of \citet{haschke2011}. We convert the color-excess values to the $E(B-V)$ values using the relation: $E(V-I) = 1.38 \times E(B-V)$ \citep{tammann2003}. We make use of the following conversion factors 
\citep{schlegel1998} to get the extinction in $V,I$ bands:

\begin{equation}
\begin{aligned}
A_V &= 3.32 \times E(B-V), \\
A_I &= 1.94 \times E(B-V).
\end{aligned}
\end{equation}

We note that the reddening law towards the central Galactic bulge is not standard \citep{popowski2000,udalski2003,nishiyama2006,nishiyama2008, nishiyama2009, nataf2013, matsunaga2013} where the optical extinction is significantly large. Therefore, we adopt two different methods to correct the optical colors of RR Lyrae and type II Cepheid variables in the Galactic bulge. In the first, we obtain the color excess $E(J-\rm{K_{s}})$ values using their galactic latitudes and longitudes within a boxsize of $2'$ from the extinction-maps of \citet{gonzalez2012}\footnote{\url{http://mill.astro.puc.cl/BEAM/calculator.php}}. The extinction maps also provide a value of the extinction in $K_s$-band, $A_{K_s} = 0.689 E(J-K_s)$ using the Cardelli extinction-law \citep{cardelli1989}. The optical total-to-selective absorption ratios used from the Cardelli extinction-law are $\frac{A_{K_s}}{A_V} = 0.114$ and $\frac{A_I}{A_V} = 0.479$. In the second method, we take color excess $E(V-I)$ and extinction $A_I$ obtained using the positions (RA/Dec) in the extinction calculator based on the OGLE data \citep{nataf2013}\footnote{\url{http://ogle.astrouw.edu.pl/cont/4_main/ext/ogleiii_gb/index.html}}. This extinction law makes use of the $E(J-K_S)$ values from \citet{gonzalez2012} and provides a relation: $A_I = 0.7465E(V-I) + 1.3700E(J-K_S)$. The $A_V$ values are estimated using the relation $E(V-I)=A_V-A_I$.
For the globular cluster M4, the color excess of $E(B-V) = 0.37\pm0.10$ is taken from \citet{hendricks2012} and the total-to-selective extinctions of $A_{3.6}/E(B-V)=0.203$ and $A_{4.5}/E(B-V)=0.156$ from \citet{monson2012} to finally get the extinction corrections of $A_{3.6}=0.075\pm0.020$ mag and $A_{4.5}=0.058\pm0.016$ mag. These extinction corrections are applied to the magnitudes and colors at minimum, mean and maximum light during the pulsation cycle. The light curve structure as well as the Fourier parameters and amplitudes are considered to be independent of the extinction correction for the present analysis. However, the interstellar reddening may affect the light curves as a second-order effect \citep{schmidt1982,mccall2004,hendricks2012}.

\begin{table*}
\caption{The light curve parameters of observed RR Lyrae stars. The columns provide the source, filter ($\lambda$), mode of pulsation, Star ID, logarithmic period, order of Fourier fit, amplitude ($A$), mean magnitude ($m_0$) and the Fourier amplitude ($R_{21}$,$R_{31}$) and phase ($\phi_{21}$,$\phi_{31}$) parameters.}
\centering
\scalebox{0.95}{
\begin{tabular}{c c c c c c c c c c c c}
\hline\hline
Source & $\lambda$ & Mode & Star ID & $\log(P)$ & order & A  & $m_0$ & $R_{21}$ & $R_{31}$ & $\phi_{21}$ & $\phi_{31}$\\
& & & & & & & $\sigma_{m_0}$ & $\sigma_{R_{21}}$ & $\sigma_{R_{31}}$ & $\sigma_{\phi_{21}}$ & $\sigma_{\phi_{31}}$\\
[0.5ex]
\hline \hline

Bulge&	V&	FU&	OGLE-BLG-RRLYR-00162&	-0.263&	5&	1.014&	17.272&	0.608&	0.799&	3.119&	5.996\\	
& & & & & & & 0.027& 0.164& 0.281& 0.392& 0.200\\
Bulge&	V&	FU&	OGLE-BLG-RRLYR-00172&	-0.321&	6&	0.976&	16.946&	0.429&	0.272&	2.127&	5.130\\	
& & & & & & & 0.033& 0.127& 0.042& 0.208& 0.691\\
...&	...&	...&	...&	...&	...&	...&	...&	...&	...&	...&	...\\
\\
LMC&	V&	FU&	OGLE-LMC-RRLYR-00579&	-0.227&	4&	0.607&	19.746&	0.372&	0.237&	2.399&	5.164\\	
& & & & & & & 0.004& 0.028& 0.027& 0.088& 0.133\\
LMC&	V&	FU&	OGLE-LMC-RRLYR-00595&	-0.214&	4&	0.704&	19.776&	0.456&	0.270&	2.390&	5.272\\	
& & & & & & & 0.003& 0.018& 0.017& 0.049& 0.079\\
...&	...&	...&	...&	...&	...&	...&	...&	...&	...&	...&	...\\
\\
SMC&	V&	FU&	OGLE-SMC-RRLYR-0001&	-0.253&	4&	1.014&	19.588&	0.338&	0.322&	2.155&	4.599\\	
 & & & & & & & 0.010& 0.040& 0.037& 0.136& 0.170\\
SMC&	V&	FU&	OGLE-SMC-RRLYR-0002&	-0.226&	4&	0.729&	19.612&	0.483&	0.273&	2.493&	5.443\\	
& & & & & & & 0.009& 0.047& 0.042& 0.106& 0.192\\
...&	...&	...&	...&	...&	...&	...&	...&	...&	...&	...&	...\\
\\
M4&	3.6&	FU&	V5&	-0.206&	4&	0.156&	10.884&	0.230&	0.081&	4.812&	3.473\\	
& & & & & & & 0.004& 0.083& 0.082& 0.367& 1.197\\
M4&	3.6&	FU&	V7&	-0.302&	4&	0.299&	11.090&	0.424&	0.257&	4.046&	1.667\\	
& & & & & & & 0.007& 0.096& 0.084& 0.278& 0.462\\
...&	...&	...&	...&	...&	...&	...&	...&	...&	...&	...&	...\\
\hline
\end{tabular}}
\begin{tablenotes}
      \small
      \item \textbf{Notes:} This table is available entirely in a machine-readable form in the online journal as supporting information.
    \end{tablenotes}
\label{tab:Observed}
\end{table*}

\section{Fourier analysis of RR Lyrae light curves}
\label{sec:fourier}

Fourier analysis method is discussed in detail, for example, in \citet{deb2009} and \citet{bhardwaj2015}. In brief, the theoretical and photometric light curve data of RR Lyrae stars are fitted with the Fourier sine-series of the form:

\begin{equation}
m(x) = m_0 + \sum_{k=1}^{N}A_k sin(2 \pi kx+\phi_k),
\label{eq:fourier}
\end{equation}

\noindent where $x$ is the pulsation phase. The periodicity in the light curves is used to fold the entire light curve into one phase cycle $(0 \leq x \leq 1) $. In equation~\ref{eq:fourier}, $m_0$ is the mean magnitude and $N$ is the order of the fit. We have used $N = 20$ for the theoretical light curves, $N = 4$ for the data from the globular cluster M4. 
For the photometric light curves from OGLE, $N$ is obtained using the Bart's criteria \citep{bart1982} by varying it from 4 to 8. To obtain the best-quality light curves, we restrict our sample to the stars for which more than 30 observations are available in the OGLE-IV catalogue.

Fourier amplitude and phase coefficients ($A_k$ and $\phi_k$) are used to determine Fourier amplitude ratios and phase differences:
\begin{equation}
\begin{aligned}
R_{k1} &= \frac{A_k}{A_1}, \\
\phi_{k1} &= \phi_k - k\phi_1,
\label{eq:params}
\end{aligned}
\end{equation}
\noindent where, $k > 1$ and $\rm 0 \leq \phi_{k1} \leq 2\pi$. The errors in the Fourier Parameters are calculated using the propagation of errors in the Fourier coefficients \citep[see][]{deb2010}.

We carried out a Fourier analysis of both theoretical and observed light curves of RR Lyrae variables. We find that the median value of standard deviation around Fourier fitted light curves does not exceed $\sim$0.06 mag in all of our samples in both $V$ and $I$ filters. The errors on the mean magnitudes and the Fourier parameters obtained from fitting the models are of the order of $10^{-4}$ and are, therefore, not considered in our analysis. Tables~\ref{tab:Models} and \ref{tab:Observed} provide light curve parameters of the theoretical and observed light curves of RR Lyrae stars, respectively. In the following subsections, we discuss the variation of their light curve parameters as a function of period, wavelength and metallicity.

\subsection{Amplitude parameters}

The pulsation models of RR Lyrae stars are able to reproduce the observed light and velocity variations and the topology of the instability strip \citep{marconi2015}. The peak-to-peak amplitude is defined as the difference between maximum and minimum of light variations:
\begin{equation}
(M_\lambda)_{amp} = (M_\lambda)_{min} - (M_\lambda)_{max},
\end{equation}
where $(M_\lambda)_{min}$ and $(M_\lambda)_{max}$ are the minimum and maximum magnitudes in the $\lambda$-band, respectively, obtained from the best-order Fourier fits. 

Fig.~\ref{fig:A-P} shows the variation of the mean theoretical amplitudes with period, wavelength and metallicity. The mean amplitudes are obtained by taking average in a bin-size of $\log(P) = 0.1$ dex, moving in steps of 0.03 dex. The error bars represent the standard deviation on the mean value within a given bin-size. We note that there is a range of M,L,T values in a given period bin and their impact on light curve parameters as a function of period will be discussed in the following subsections. We observe a decrease in amplitude with an increase in wavelength. This is similar to the results for Cepheid variables \citep{bhardwaj2017} and this is expected because the effective temperature of an RR Lyrae star is such that the Planck function peaks at visible wavelengths. Thus, the temperature dependence of visible flux scales as $R^{2}T_{e}^4$. On the other hand, since the infrared wavelengths lie on the Rayleigh-Jeans tail of the Planck function, the temperature dependence of $K$-band flux scales as $R^{2}T_{e}^{1.6}$ \citep{jameson1986, smith2015}. At longer wavelengths, $K$-band and $L$-band amplitudes are similar. We have separated all the models in the three metal-abundance ranges around Z=0.004 for plotting but we note that these results hold for all sets of models with a fixed metal-abundance.

However, for some short-period models ($\log(P)<-0.1$ for Z=0.004 and $\log(P)<0.1$ for Z$<$0.004), the $U$-band mean amplitudes are lower than the mean amplitudes in $B$-band. This depends on the effective temperature of the RR Lyrae star. If the effective temperature is such that the Planck function peaks at wavelength corresponding to the $B$-band, $U$-band now lies at the shorter-wavelength side of the peak and thus, $B_{amp}$ is the largest. The wavelengths on either side of the peak have amplitudes less than $B_{amp}$. In contrast, if the effective temperature is such that the Planck function peaks at $U$-band, all other bands will have amplitudes less than $U_{amp}$. This can also be seen from the results listed in Table 6 of \citet{monson2017} for Galactic field RR Lyrae variables.

We also note the variation of the mean amplitudes with periods and find a decrease in amplitude with increase in period for both Z$>$0.004 and Z=0.004, with a minimum at around $\log(P)=0$. For Z$>$0.004, this is followed by an increase in the mean amplitude with period and a maximum occurs at around $\log(P)=0.2$. For the case of Z$<$0.004, the monotonic decrease in the mean amplitude with period is followed by a break in the pattern around $\log(P)=-0.1$. These changes in the mean amplitudes with period are more distinct in the optical bands as compared to those in the infrared bands.

\begin{figure*}
\centering
\includegraphics[width=1.0\textwidth,keepaspectratio]{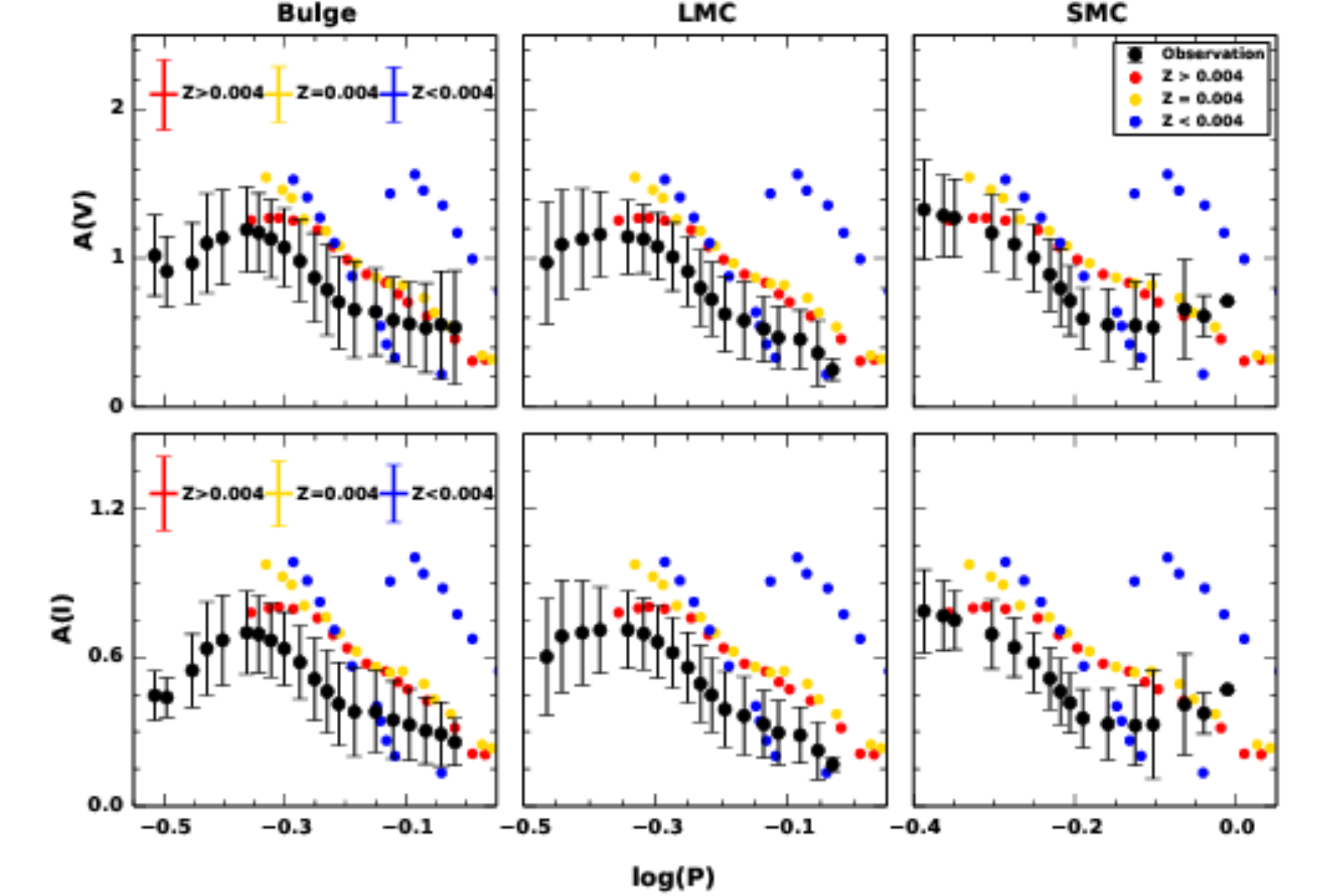}
\caption{A comparison of the mean amplitudes for RRab models with the mean observed amplitudes for RRab stars in Bulge, LMC and SMC in the $V$ and $I$- bands. The representative error bars are plotted for both $VI$ bands in the first-column.}
\label{fig:meanA-P}
\end{figure*}

The comparison of the mean amplitudes from the theoretical RRab models with the mean observed amplitudes for RRab stars in the Bulge, LMC and SMC at optical wavelengths ($VI$) is depicted in Fig. \ref{fig:meanA-P}. The mean amplitudes for models with Z$>$0.004 and Z=0.004 have similar trend to RRab stars in the Bulge, LMC and SMC, with a monotonic decrease in their mean amplitudes with period in the range $-0.35<\log(P)<0$. The mean amplitudes for all Z with $\log(P)<-0.1$ match well with the observed amplitudes in the $VI$-band, given the large uncertainties. However, the longer period models ($\log(P)>-0.1$) for Z$<$0.004 display significantly large mean amplitudes when compared with the observations. The amplitude ranges are also consistent between the models and the observations, with marginally better consistency in the $V$-band than in the $I$-band. In general, the mean amplitude values from the models are slightly higher than those from the observations. \citet{criscienzo2004b} had suggested that an increase in the mixing length parameter can cause a decrease in the pulsation amplitudes while keeping the light curve structure unchanged for RR Lyrae stars - this is similar to the results for Cepheid variables \citep{fiorentino2007, bhardwaj2017}. We note that the pulsation amplitudes of theoretical light curves are affected by the uncertainties on the assumed convective efficiency \citep[see, e.g.,][]{fiorentino2007}. 

\subsection{Theoretical Fourier Parameters}

\begin{figure}
\centering
\includegraphics[width=1.0\columnwidth,keepaspectratio]{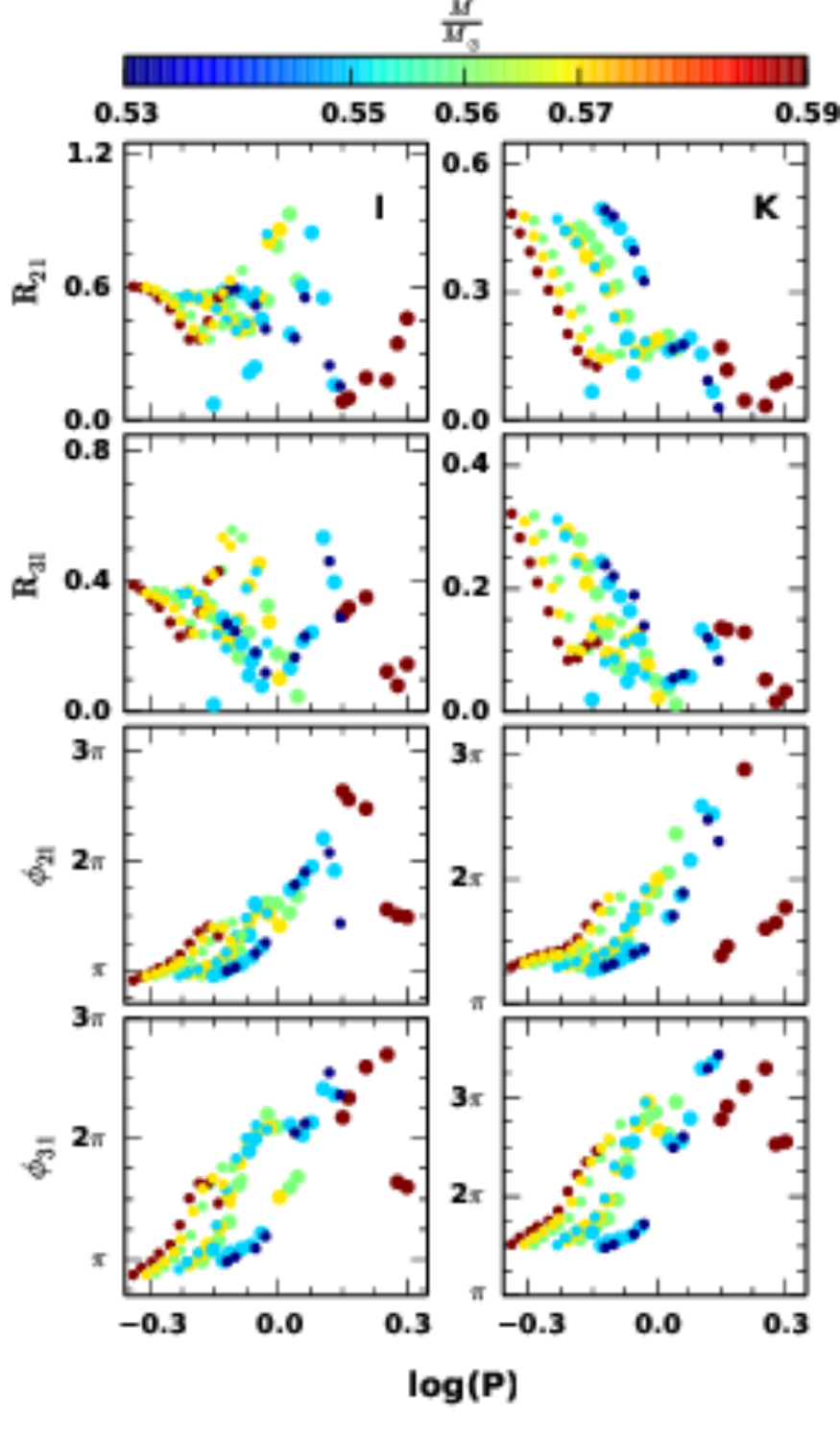}
\caption{Fourier parameters for theoretical light curves of RRab stars in the $I$ and $K$-bands for Z=0.004. The colorbar represents the different stellar mass models. For a given mass, small/large symbols represent low/high luminosity values as listed in Table~\ref{tab:Data}.}
\label{fig:Fourier_mass}
\end{figure}

We present the variation of $I$ and $K$-band Fourier parameters as a function of mass for Z=0.004 in Fig.~\ref{fig:Fourier_mass}. Note that some of the Fourier phase parameters ($\phi_{21}$ and $\phi_{31}$) have been offset by 2$\pi$ for visualisation purposes, for this figure as well as the following figures involving phase parameters in this paper. A smooth grid of models with a range of masses and luminosities enables us to predict how mass affects the Fourier parameters. The various sets of M-L pairs are clearly separated on the Fourier amplitude plane as a function of period in the $K$-band. However, this is not seen for models with $P > 1$d and also at optical wavelengths. In $K$-band, the separation of different M-L pairs is also distinct and clear on the phase parameter plane and we also note that the two different luminosity values for a given mass are well-separated. Further, for a fixed pair of M-L, the amplitude parameters decrease and the phase parameters increase with increase in mass in the overlapping period range. The near-infrared data available at present either consists of single-epoch observations or a few tens of observations. These sparsely sampled light curves do not provide accurate and precise values of Fourier parameters. With well-sampled near-infrared time-series data becoming available in near-future, such comparison with models can be used to provide constraints on the M-L combinations of RR Lyrae variables.

\begin{figure*}
\centering
\includegraphics[width=1.0\textwidth,keepaspectratio]{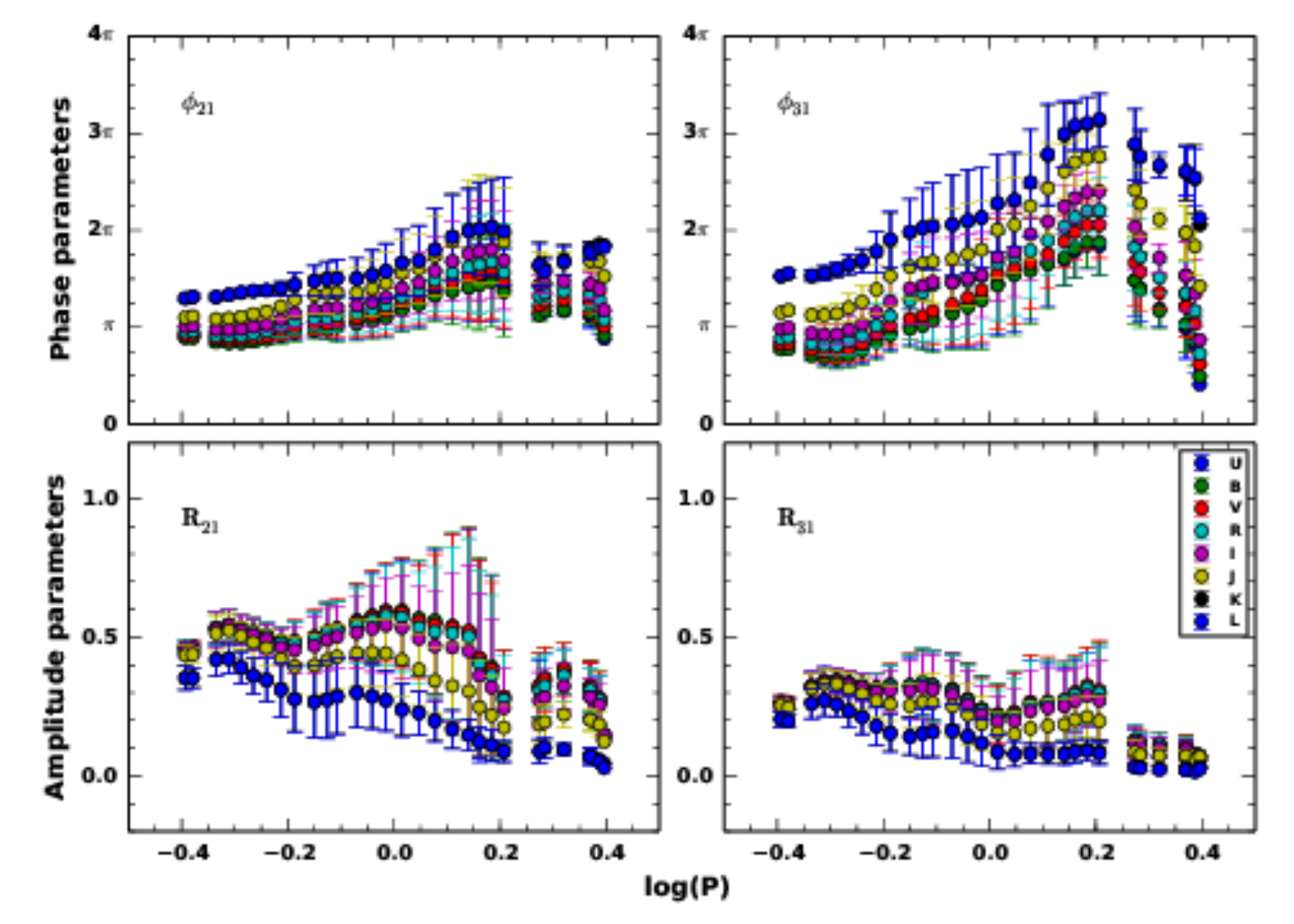}
\caption{Multi-wavelength mean Fourier parameters for RRab models with the different compositions combined.}
\label{fig:Fourier_allbands}
\end{figure*}

The variation of the mean Fourier amplitude ($R_{21}$ and $R_{31}$) and phase ($\phi_{21}$ and $\phi_{31}$) parameters with wavelength is shown in Fig.~\ref{fig:Fourier_allbands} for all the models of different Z values combined together. We observe a distinct and clear trend in the mean Fourier parameters as a function of wavelength. For a given period, the mean amplitude parameters decrease with increase in wavelength and the mean phase parameters increase with wavelength. Similar variation was also observed for the Cepheid variables with both theoretical and observed data \citep{bhardwaj2017}. The optical mean amplitude parameters display greater scatter when compared to infrared bands as the metallicity effects are more pronounced at shorter wavelengths \citep{longmore1986, bono2001, sollima2006}. Further, the scatter in the mean phase parameters also increases for $P > 1$d. 

\begin{figure}
\centering
\includegraphics[width=1.0\columnwidth,keepaspectratio]{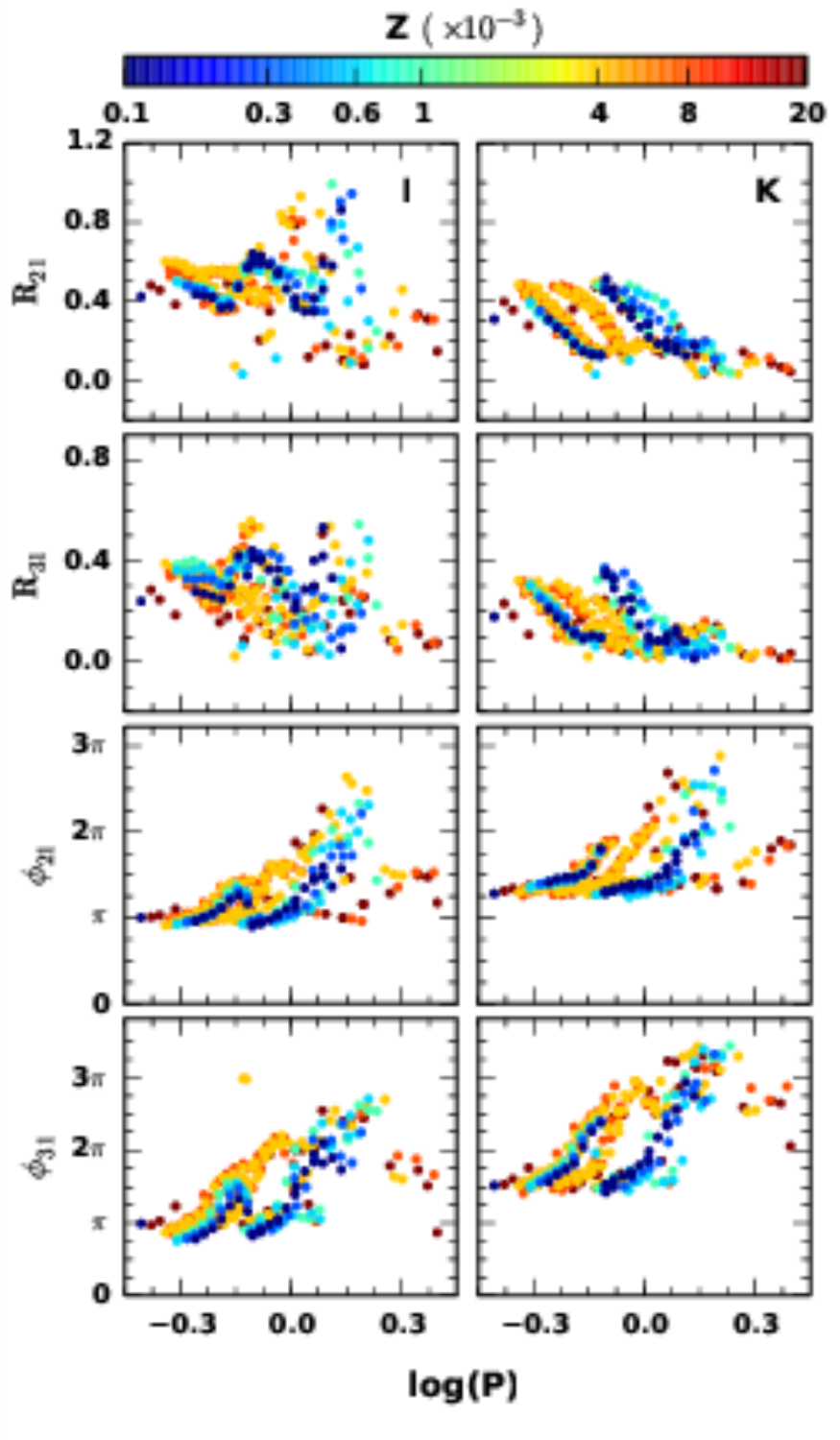}
\caption{A comparison of the theoretical Fourier parameters for RRab models over the $I$ and $K$-bands.}
\label{fig:Fourier_diffZ}
\end{figure}

Fig.~\ref{fig:Fourier_diffZ} displays the variation of the Fourier parameters with period as a function of different compositions at $IK$ wavelengths. 
We find that for the short period RR Lyrae stars ($\log(P) < -0.15$), the values of $R_{21}$ are greater for metal-rich models at optical bands while no clear trend is seen in $R_{31}$. We also observe separation of RR Lyrae models based on different sequences of metal-abundances in $K$-band amplitude parameter plane. As expected, the phase parameters display a better correlation with metal-abundance both at optical and infrared bands. The phase parameters increase with metal-abundance at a given period in optical bands while in $K$-band the variation is similar for a fixed set of M-L pair. Note that the correlation of Fourier phase parameters with metallicity has resulted in several empirical $P-[Fe/H]-\phi_{31}$ relations in the literature \citep{jurcsik1996, smolec2005,morgan2007, jurcsik2009, nemec2013, ngeow2016}.

We also carry out a quantitative analysis of the difference in the mean Fourier parameters (obtained by binning the entire period range in steps of log($P$)=0.15 dex) across two filters $V$ and $I$ for the cases $-$ Z=0.02, Z=0.008 and Z=0.004. We do not find any significant difference within 3$\sigma$ uncertainties in the mean Fourier parameters, except for the $R_{21}$ value in the period bin 0.30-0.45.

\subsection{Comparison of the theoretical and observed Fourier parameters} \label{sec:fourier_obsthr}

\begin{figure*}
\centering
\includegraphics[width=1.0\textwidth,scale=1]{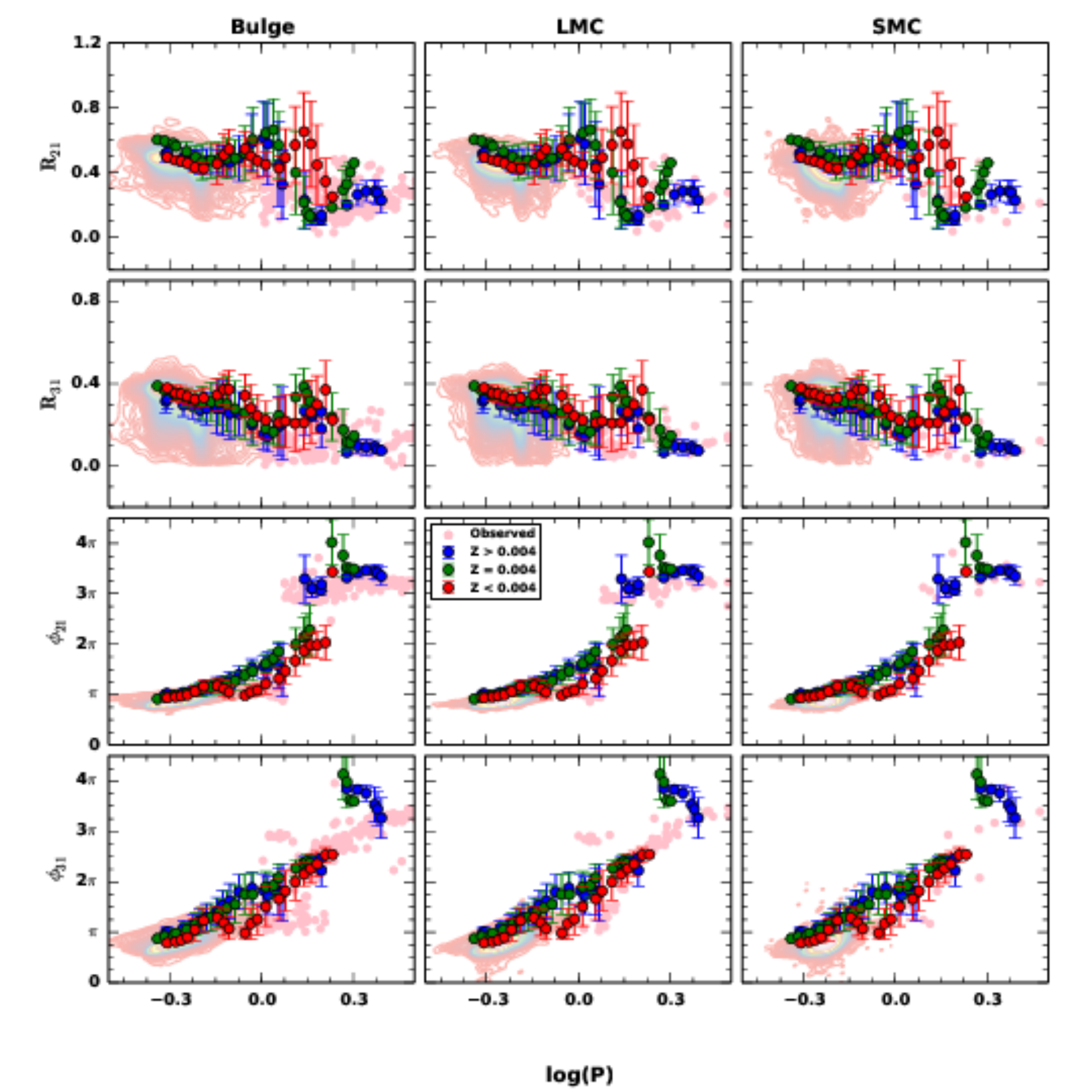}
\caption{A comparison of the theoretical mean Fourier parameters for RRab stars with compositions Z $>$ 0.004, Z = 0.004 and Z $<$ 0.004 with the observed mean Fourier parameters from OGLE-IV RRab stars (shown in contour lines) and type II Cepheids (shown as scatter points) in the Bulge, LMC and SMC from the OGLE catalog in the $I$-band.}
\label{fig:Fourier}
\end{figure*}

We present a comparison of the theoretical mean Fourier parameters for RRab stars having metal-abundances, Z $>$ 0.004, Z = 0.004 and Z $<$ 0.004, with the observed $I$-band Fourier parameters of RRab and type II Cepheid variables in the Bulge, LMC and SMC  from the OGLE catalog in Fig.~\ref{fig:Fourier}. We find that the Fourier parameters from the models are consistent with those from observations for all the cases over most of the period bins ($P<1$d), given the dispersion in observed parameters. However for $P > 1$, we compare the theoretical amplitude parameters with the observed parameters for type II Cepheids and find that average values are consistent within errors, although the $R_{21}$ values for $0<\log(P)<0.15$ for in case of $Z<0.004$ show a greater offset. In case of $\phi_{21}$, there is a discrepancy between theoretical and observed parameters in the period bin of $0<\log(P)<0.2$. However, $\phi_{31}$ values match well with observations over the all period ranges. While there is an overall consistency between theoretical and observed light curve parameters, some of these discrepancies related to the amplitude parameters can be resolved by adopting a higher mixing length (see e.g. \citealt{marconi2005,marconi2007} for RR Lyraes and \citealt{marconi2013,marconi2013b} for classical Cepheids) or with changes in opacity \citep{kanbur2018}. The discrepancy at longer periods corresponding to redder stars may be resolved by assuming different mixing length parameters in the blue and in the red parts of the HR diagram when modelling RR Lyrae pulsation \citep[see e.g.][]{criscienzo2004b}. We emphasize here that it is better to use one set of mixing length/turbulent convection parameters or artificial viscosity parameters for the whole set of RR Lyraes unless evidence indicates a specific variation with period in these parameters. We also compared limited NIR data available in the globular clusters for RRab stars \citep{neeley2015} and find that the observed Fourier parameters are consistent with models at longer wavelengths. However, the poor phase coverage in NIR leads to a greater dispersion in the Fourier plane and additional data with well-sampled light curves may lead to a more rigorous comparison, thus, constraining the model parameters \citep{marconi2017}.

\begin{table*}
\caption{RRab stars in the LMC with period and Fourier coefficients that match with the theoretical models. The first three columns are from observations while the last six columns are from the models. The extinction-corrected apparent magnitude ($m_0$) and absolute magnitude $M$ are used to estimate the distance modulus $\mu$ to the LMC.}

\centering
\scalebox{0.9}{
\begin{tabular}{c c c c c c c c c c c}
\hline\hline
Star ID & $\log(P)$ & $m_0$ & $M$ & $\mu$ & Z & Y &  $\frac{M}{M_{\odot}}$ & $\log\frac{L}{L_{\odot}}$ & $\log\frac{R}{R_{\odot}}$ & $T_e$	\\
[0.5ex]
\hline \hline
OGLE-LMC-RRLYR-34041	&	-0.193	&	18.483$\pm$0.006	&	0.059$\pm$0.0	&	18.424$\pm$0.006	&	0.0200	&	0.270	&	0.51	&	1.69	&	0.702	&	6800\\
OGLE-LMC-RRLYR-25897	&	-0.277	&	18.358$\pm$0.011	&	0.316$\pm$0.001	&	18.043$\pm$0.011	&	0.0080	&	0.256	&	0.57	&	1.58	&	0.676	&	6600\\
OGLE-LMC-RRLYR-25411	&	-0.307	&	18.959$\pm$0.014	&	0.3$\pm$0.001	&	18.659$\pm$0.014	&	0.0040	&	0.250	&	0.59	&	1.61	&	0.689	&	6600\\
OGLE-LMC-RRLYR-29312	&	-0.307	&	18.955$\pm$0.009	&	0.3$\pm$0.001	&	18.655$\pm$0.009	&	0.0040	&	0.250	&	0.59	&	1.61	&	0.689	&	6600\\
OGLE-LMC-RRLYR-25897	&	-0.277	&	18.358$\pm$0.011	&	0.276$\pm$0.001	&	18.082$\pm$0.011	&	0.0040	&	0.250	&	0.59	&	1.61	&	0.702	&	6500\\
OGLE-LMC-RRLYR-38981	&	-0.249	&	18.845$\pm$0.006	&	0.245$\pm$0.001	&	18.601$\pm$0.006	&	0.0040	&	0.250	&	0.59	&	1.61	&	0.716	&	6400\\
OGLE-LMC-RRLYR-25287	&	-0.231	&	18.398$\pm$0.012	&	0.221$\pm$0.001	&	18.177$\pm$0.012	&	0.0040	&	0.250	&	0.59	&	1.61	&	0.730	&	6300\\
OGLE-LMC-RRLYR-25897	&	-0.277	&	18.358$\pm$0.011	&	0.248$\pm$0.001	&	18.11$\pm$0.011	&	0.0040	&	0.250	&	0.57	&	1.63	&	0.698	&	6600\\
OGLE-LMC-RRLYR-21626	&	-0.236	&	18.593$\pm$0.007	&	0.131$\pm$0.001	&	18.462$\pm$0.007	&	0.0010	&	0.245	&	0.64	&	1.67	&	0.735	&	6500\\
OGLE-LMC-RRLYR-26471	&	-0.238	&	18.889$\pm$0.008	&	0.131$\pm$0.001	&	18.758$\pm$0.008	&	0.0010	&	0.245	&	0.64	&	1.67	&	0.735	&	6500\\
OGLE-LMC-RRLYR-31944	&	-0.234	&	18.701$\pm$0.005	&	0.131$\pm$0.001	&	18.57$\pm$0.005	&	0.0010	&	0.245	&	0.64	&	1.67	&	0.735	&	6500\\
OGLE-LMC-RRLYR-32942	&	-0.237	&	18.842$\pm$0.006	&	0.131$\pm$0.001	&	18.712$\pm$0.007	&	0.0010	&	0.245	&	0.64	&	1.67	&	0.735	&	6500\\
OGLE-LMC-RRLYR-33913	&	-0.251	&	18.72$\pm$0.005	&	0.131$\pm$0.001	&	18.589$\pm$0.005	&	0.0010	&	0.245	&	0.64	&	1.67	&	0.735	&	6500\\
OGLE-LMC-RRLYR-36521	&	-0.250	&	18.607$\pm$0.007	&	0.131$\pm$0.001	&	18.477$\pm$0.007	&	0.0010	&	0.245	&	0.64	&	1.67	&	0.735	&	6500\\
OGLE-LMC-RRLYR-38007	&	-0.242	&	18.619$\pm$0.006	&	0.131$\pm$0.001	&	18.489$\pm$0.006	&	0.0010	&	0.245	&	0.64	&	1.67	&	0.735	&	6500\\
OGLE-LMC-RRLYR-39036	&	-0.271	&	17.479$\pm$0.004	&	0.122$\pm$0.001	&	17.357$\pm$0.004	&	0.0006	&	0.245	&	0.67	&	1.69	&	0.731	&	6600\\
OGLE-LMC-RRLYR-29103	&	-0.237	&	18.846$\pm$0.005	&	0.092$\pm$0.001	&	18.754$\pm$0.005	&	0.0006	&	0.245	&	0.67	&	1.69	&	0.745	&	6500\\
OGLE-LMC-RRLYR-30469	&	-0.235	&	18.676$\pm$0.006	&	0.092$\pm$0.001	&	18.584$\pm$0.006	&	0.0006	&	0.245	&	0.67	&	1.69	&	0.745	&	6500\\
OGLE-LMC-RRLYR-31083	&	-0.250	&	18.763$\pm$0.007	&	0.092$\pm$0.001	&	18.671$\pm$0.007	&	0.0006	&	0.245	&	0.67	&	1.69	&	0.745	&	6500\\
OGLE-LMC-RRLYR-35018	&	-0.246	&	18.81$\pm$0.006	&	0.092$\pm$0.001	&	18.718$\pm$0.006	&	0.0006	&	0.245	&	0.67	&	1.69	&	0.745	&	6500\\
OGLE-LMC-RRLYR-26708	&	-0.278	&	19.089$\pm$0.012	&	0.085$\pm$0.001	&	19.004$\pm$0.012	&	0.0003	&	0.245	&	0.72	&	1.72	&	0.733	&	6700\\
OGLE-LMC-RRLYR-25600	&	-0.231	&	18.642$\pm$0.007	&	0.025$\pm$0.001	&	18.616$\pm$0.007	&	0.0003	&	0.245	&	0.72	&	1.72	&	0.760	&	6500\\
OGLE-LMC-RRLYR-28072	&	-0.252	&	18.521$\pm$0.01	&	-0.034$\pm$0.001	&	18.556$\pm$0.01	&	0.0001	&	0.245	&	0.80	&	1.76	&	0.766	&	6600\\
OGLE-LMC-RRLYR-25600	&	-0.231	&	18.642$\pm$0.007	&	-0.063$\pm$0.001	&	18.704$\pm$0.007	&	0.0001	&	0.245	&	0.80	&	1.76	&	0.779	&	6500\\
OGLE-LMC-RRLYR-10615	&	-0.206	&	18.791$\pm$0.009	&	-0.089$\pm$0.001	&	18.88$\pm$0.009	&	0.0001	&	0.245	&	0.80	&	1.76	&	0.793	&	6400\\
\hline
\end{tabular}}
\label{tab:similar_lc}
\end{table*}

\begin{figure*}
\centering
\includegraphics[width=1.0\textwidth,scale=1]{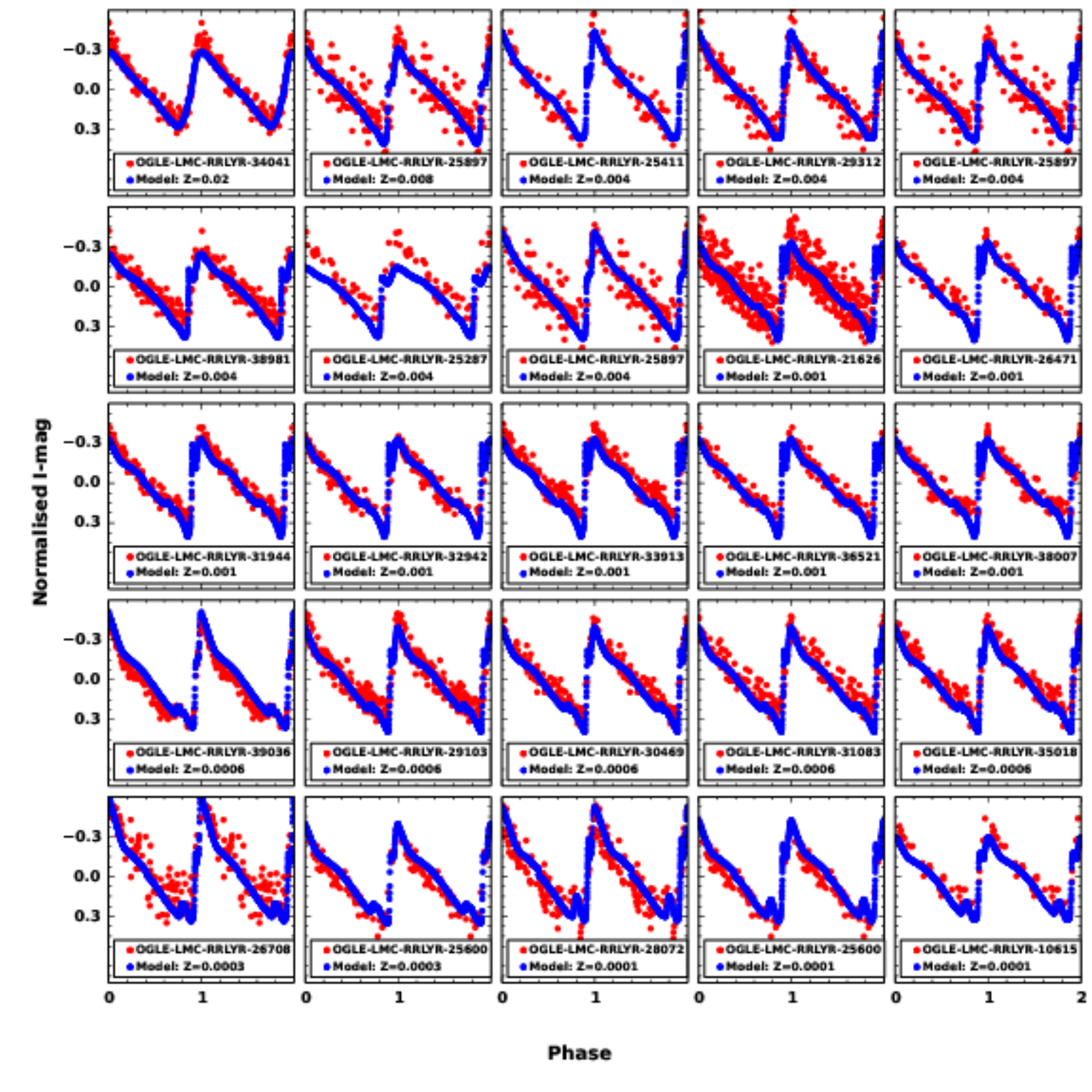}
\caption{The light curves of the 25 stars in the LMC (in red) that match with models (in blue). The period of these stars and the properties of the corresponding models such as the stellar mass, luminosity, radius and effective temperature may be found in Table~\ref{tab:similar_lc}.}
\label{fig:similarFP_lc}
\end{figure*}

\begin{figure*}
\centering
\includegraphics[width=1.0\textwidth,scale=1]{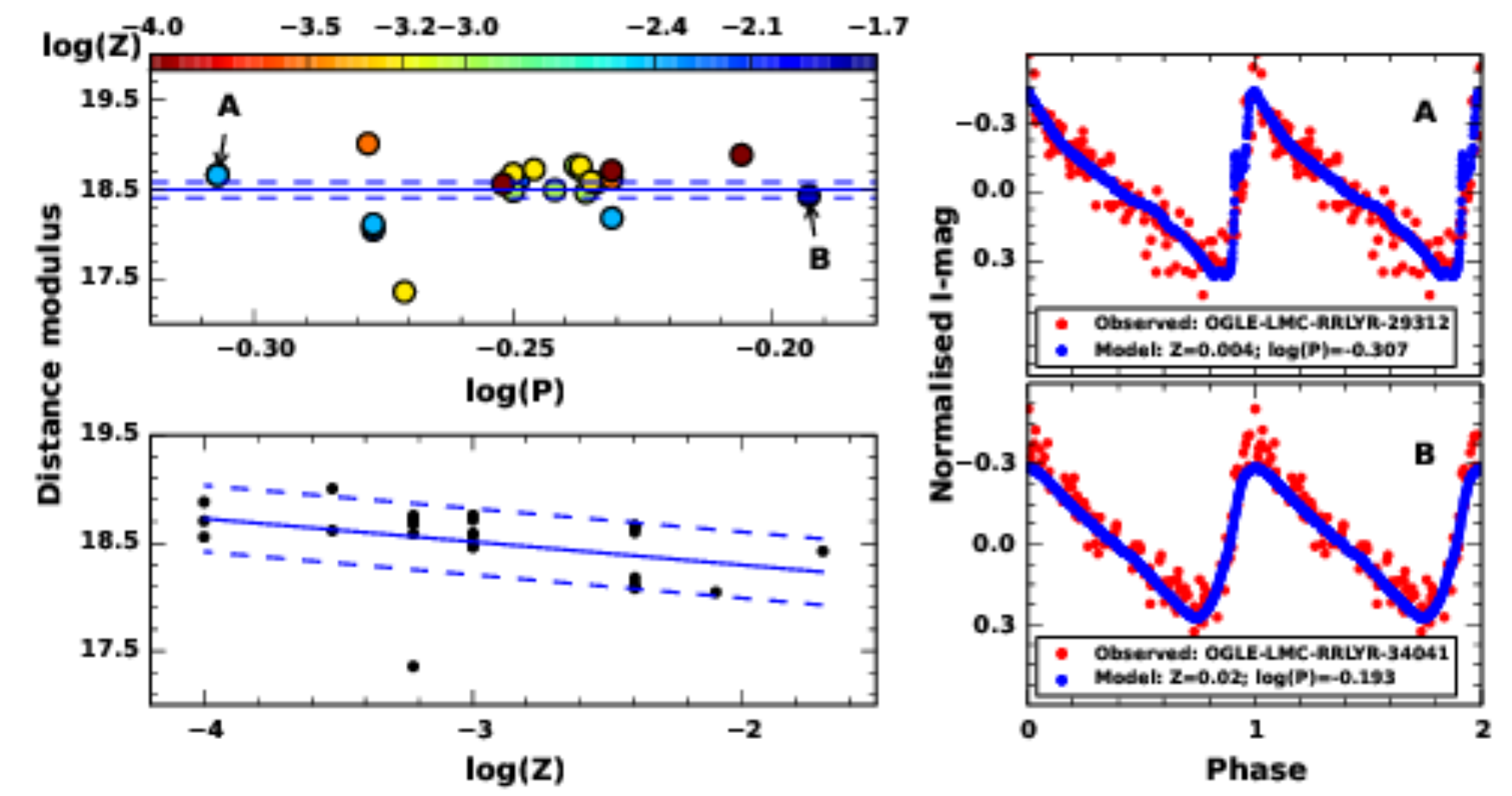}
\caption{The left panels display the variation of distance modulus as a function of period (top) and metal-abundance (bottom). Solid lines represent $\mu_\textrm{LMC} = 18.49 \pm 0.09$ in the top panel and best fit linear regression in the bottom panel while the dashed lines represent 1$\sigma$ error in each case. The right panel shows theoretical and observed (normalised with respect to median magnitude) $I$-band light curves of two RR Lyraes that are consistent on the Fourier plane.}
\label{fig:similar_lc}
\end{figure*}

One direct consequence of the comparative light curve analysis is that we can get absolute magnitude for an observed RR Lyrae star, provided the model fits the observed light curve. This will allow us to obtain robust distance estimates using theoretical models. It is known that the RR Lyrae stars do not obey a well-defined $PLR$ at optical wavelengths but exhibit a strong $PLR$ in the infrared bands \citep{longmore1986, bono2001, catelan2004, sollima2006, muraveva2015, braga2015, neeley2015}. Despite the recent progress, the absolute calibration of RR Lyrae $PLR$ is not well-constrained, mainly due to the lack of parallax measurements. Theoretical $PLZ$ relations for RR Lyrae stars can be used as calibrators for the extragalactic distance scale but the contribution due to the metallicity is not well-understood as there is discrepancy between the theoretical and observed $PLZ$ relations in the literature. However, we know that the lower order Fourier coefficients ($A_i$ and $\phi_i$) provide a good first order measure of the light curve structure. Therefore, we select similar $I$-band light curves in the theoretical models and in the observations from the OGLE-IV data. We select good-quality light curves by imposing the following conditions:  

\begin{enumerate}
\item $\left| \log(P)_{\mathrm{model}}-\log(P)_{\mathrm{observed}} \right| \leq 0.01$,
\item $\left| A_{i,\mathrm{model}} - A_{i,\mathrm{observed}} \right| \leq 3\sigma_{A_{i,\mathrm{observed}}}$,
\item $\left| \phi_{i,\mathrm{model}}- \phi_{i,\mathrm{observed}} \right| \leq 3\sigma_{\phi_{i,\mathrm{observed}}}$,
\end{enumerate}
where $1 \leq i \leq 3$. We find 25 RR Lyrae stars in the LMC that pass these criteria and have light curves that match with theoretical models $-$ the light curves are presented in Fig~\ref{fig:similarFP_lc} and the results are summarised in Table~\ref{tab:similar_lc}. We adopt the absolute magnitudes from models and extinction-corrected apparent magnitudes to obtain the distance modulus to the LMC. The average distance modulus to the LMC obtained from these 25 stars is found to be $\mu_\textrm{LMC}=18.51\pm0.07$, in a very good agreement with the published result, $18.49\pm0.09$ from \citet{grijs2014} based on several tracers and other primary distance indicators, e.g., late-type eclipsing binaries \citep[{18.493 mag $\pm$ 0.008(statistical) $\pm$ 0.047 (systematic),}][]{pietrzynski2013} and Cepheids \citep[{18.47 mag $\pm$ 0.07(statistical),}][]{bhardwaj2016}. The upper left panel of Fig.~\ref{fig:similar_lc} displays the estimated distance modulus as a function of period and metal-abundance. The lower left panel plots distance modulus as a function of metallicity. The estimated distance moduli display a correlation with metal abundance (Slope = $-0.215\pm0.013$), albeit a greater scatter ($\sigma=0.307$). Right panels show the theoretical models overplotted on the normalised light curves of LMC RRab stars. The models fit the observations very well and this similarity allows a good estimate of distance modulus without accounting for any metallicity effects. We also test the robustness of distance estimate as we adopt looser constraints on matching the light curve structure. From Table~\ref{tab:similar_lc}, we see that one model may correspond to multiple stars in the observed sample. Therefore, we adopt an additional condition that only one star may correspond to one model. This allows us to compare matches that are independent of duplicates as we go to higher sigma threshold. We find that the distance to the LMC changes from 18.48 to 18.47 with 2-10 sigma threshold and the statistical uncertainties vary from 0.16 to 0.04. We note that the LMC distance does not vary significantly with looser constraints as there are more number of matches, also leading to smaller statistical uncertainties. This suggests that period is indeed the primary parameter, and even though the light curve morphology is not strictly consistent, the reasonable distances can be estimated. However, in the era of percent-level precision, it is important to consider second-order effects i.e. the contribution from M,L,T that goes into the light curve structure for a fixed composition. A denser and smoother grid will also affect this result in a positive way and show the importance of light curve structure as a secondary way to constrain models over and above period in the era of precision cosmology. Also given the consistency of models with observed light curves, we can also provide a reasonable estimate of the physical parameters of the observed stars such as the chemical composition, stellar mass, luminosity, radius and effective temperature. In order to provide robust measurement of the physical parameters, a smoother and denser grid of models is required that will be used together with an automated non-linear optimization method \citep{bellinger2016} in a future study.

\section{Period-color and amplitude-color relations}
\label{sec:PCAC}

The colors at minimum and maximum light for RR Lyrae and Cepheid variables are used to probe the interactions of stellar photosphere and hydrogen ionization front \citep{kanbur2004, bhardwaj2014}. We use theoretical models and OGLE-IV light curve data for RR Lyrae stars in the Galactic bulge, LMC and SMC to study their PC and AC relations. The colors at maximum and minimum light are defined as:

\begin{equation}
\begin{aligned}
(m_{\lambda_1} - m_{\lambda_2})_{max} &= (m_{\lambda_1})_{max} - (m_{\lambda_2})_{phmax},\\
(m_{\lambda_1} - m_{\lambda_2})_{min} &= (m_{\lambda_1})_{min} - (m_{\lambda_2})_{phmin},
\end{aligned}
\end{equation}

\noindent where $\lambda_1 < \lambda_2$ and $m$ is the apparent/absolute magnitude in case of observations/models at a particular wavelength. $(m_{\lambda_2})_{phmax}$ \& $(m_{\lambda_2})_{phmin}$ correspond to the magnitude in $\lambda_2$ at the same phase as that of $(m_{\lambda_1})_{max}$ \& $(m_{\lambda_1})_{min}$, respectively.

\subsection{Observed colors of RR Lyrae stars at maximum and minimum light}

\begin{figure*}
\centering
\includegraphics[width=1.0\textwidth,scale=1,trim = 0 0 0 0, clip]{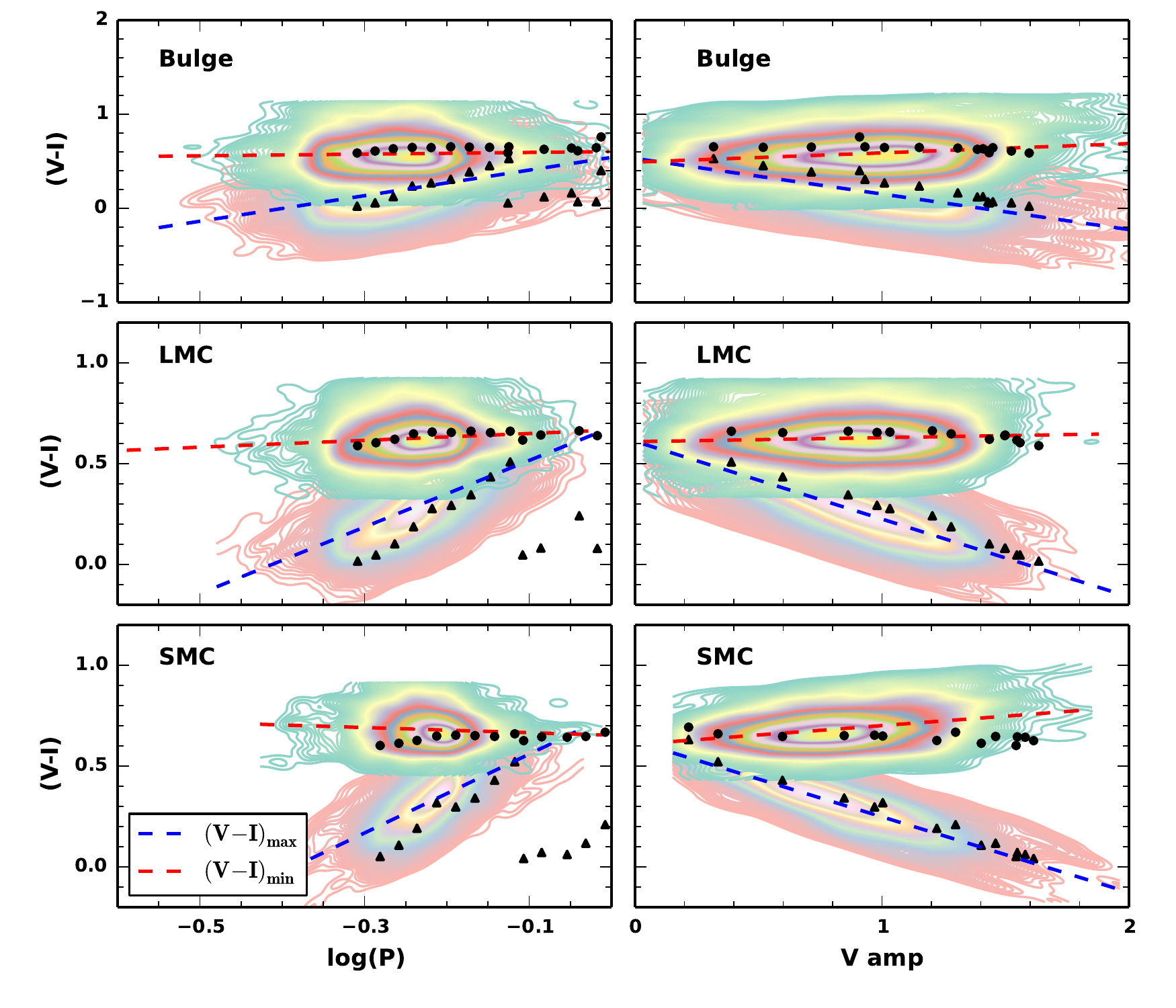}
\caption{PC and AC relations for RRab stars at minimum and maximum light for Bulge, LMC and SMC shown as contour lines with the dashed lines representing the best fit linear regressions. Black dots denote corresponding models (Z=0.001/Bulge, Z=0.00006/LMC and Z=0.0003/SMC) at minimum light while triangles represent the models at maximum light.}
\label{fig:PCAC_rrab}
\end{figure*}

\begin{figure*}
\centering
\includegraphics[width=1.0\textwidth,scale=1,trim = 0 0 0 0, clip]{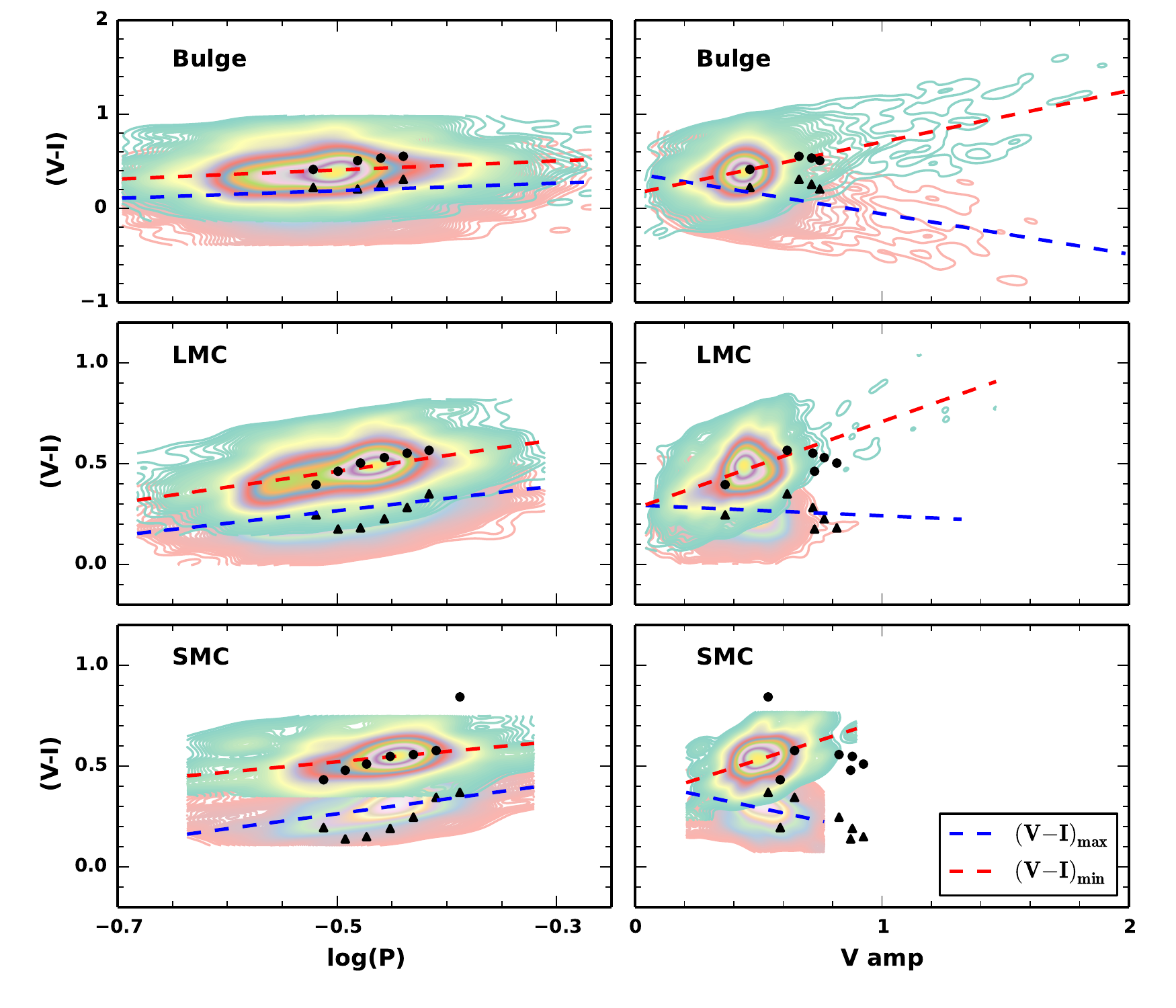}
\caption{Same as Fig.\ref{fig:PCAC_rrab} but for RRc stars.}
\label{fig:PCAC_rrc}
\end{figure*}

Figs.~\ref{fig:PCAC_rrab} and \ref{fig:PCAC_rrc} show the PC and AC relations for the Bulge, LMC and SMC RRab and RRc stars, respectively, at the phases of maximum and minimum light. We correct the observed colors for extinction as discussed in Section~\ref{sec:data} and fit linear regression to obtain robust estimates of slopes and zero-points after recursively removing 3$\sigma$ outliers in PC and AC relations. The results are tabulated in Table~\ref{tab:PCAC}.

We note that \citet{bhardwaj2014} analysed PC and AC relations for Cepheid and RR Lyrae variables in the LMC and SMC using OGLE-III data. However, the OGLE-IV light curves have more number of data-points, specially in the $V$-band and, therefore, provide an improved color estimate at minimum and maximum light. Further, the number of RR Lyrae stars have also increased significantly in the OGLE catalog of variable stars. In this work, we have also included Bulge RR Lyrae stars by accounting for the extinction using two different methods. From Table~\ref{tab:PCAC}, we find that the Galactic Bulge RRab stars have a nearly-flat or shallow PC slope at minimum light that is consistent with previous results \citep{simon1993,kanbur1995,bhardwaj2014,ngeow2017}. The corresponding AC relation is also very shallow whereas there is significant slope at maximum light for PC and AC relations. At minimum light, the slope of the PC relation is negative/positive for the RRab in the SMC/LMC while it is nearly flat for the Bulge RRab stars. At maximum light, the slope of PC relation increases going from the Bulge (1.359$\pm$0.028/1.397$\pm$0.028), LMC (1.651$\pm$0.017) to SMC (1.956$\pm$0.039), thus suggesting a possible correlation with metallicity such that higher metal-abundance leads to shallower PC$_{\mathrm{max}}$ relation. We also note that the results of the PC slopes at different phases are preserved for the Bulge RR Lyrae stars using the two different extinction methods of \citet{cardelli1989} and \citet{nataf2013}, discussed in Section~\ref{sec:data}. We compare the PC$_{\mathrm{min}}$ relations for the common RRab stars in the OGLE-III and OGLE-IV and find that these changes in slopes are preserved and, therefore, deserve further investigation. It is worth emphasizing that the PC$_{\mathrm{min}}$ relation for new RRab stars in the OGLE-IV displays a flat relation similar to OGLE-III RRab stars.

From the viewpoint of the HIF-stellar photosphere interaction as discussed in \citet{simon1993, kanbur1995, kanbur1996, bhardwaj2014, ngeow2017}, we note that these OGLE IV results are still consistent with those ideas because as Table~\ref{tab:PCAC} demonstrates, the change in slope from minimum to maximum light is very significant. At minimum light, the photosphere and HIF are engaged and so the temperature of the photosphere and hence the color at minimum light is not strongly dependent on global stellar parameters: hence a shallow PC relation slope. As the star brightens from minimum light, the temperature increases, but the HIF and stellar photosphere are still engaged \citep{kanbur1996}. In this temperature range, the temperature at which hydrogen ionizes becomes much more strongly dependent on the global stellar properties. This leads to a greater dependence on period for the PC relation at maximum light as opposed to minimum light. Following the work of \citet{simon1993} and \citet{kanbur1996}, if the PC relation is shallower at min/max, the AC relation will be steeper at max/min. We find that the AC relations in the present analysis are also completely consistent with these predictions based on the Stefan-Boltzmann law \citep[see,][for more details]{bhardwaj2014}. Here, we emphasize that unlike RRab stars, the RRc stars do not show the large difference in slopes in the PC relation between max/min, as is depicted in Fig.\ref{fig:PCAC_rrc}. This is consistent with the theory of the HIF-stellar photosphere interaction. The HIF and stellar photosphere are still engaged, but since the overtones are hotter, the engagement occurs at a higher temperature which is in a range where Saha ionization equilibrium is more sensitive to temperature.

In addition, we note that the PC$_{\mathrm{min}}$ is small but negative/positive for the SMC/LMC and the SMC has lowest metallicity of the three galaxies considered. It is worth mentioning here that while using the Bart's criteria to determine the optimum order of fit for individual stars results in a small negative PC$_{\mathrm{min}}$ slope of $-0.128\pm0.033$ for the RRab stars in SMC, using the order of fit, $N$=4 for all stars, results in a flatter PC$_{\mathrm{min}}$ slope of $-0.090\pm0.030$. Increasing the order of the fit for all stars simultaneously leads to over-fitting. Further investigation is required to determine if the PC$_{\mathrm{min}}$ slope for SMC RRab stars is indeed significantly negative. In a future study, we plan to combine these results with evolutionary models of RRab stars appropriate for the Bulge, LMC and SMC to study how the outer envelope structure and relative location of the stellar photosphere/HIF change due to metallicity and effect the slope of the PC$_{\mathrm{min}}$ relation. The minimum light color of RR Lyrae has been used to estimate reddening but a large dispersion in $(V-I)_\mathrm{min}$ suggests that caution should be used when using the properties of RR Lyrae stars at minimum light to determine reddening \citep{guldenschuh2005}. We compare here the size of the scatter in different colors at minimum light $-$  it varies from $\sim$0.02 in $r-i$ to $\sim$0.07 in $u-g$ for RRab in SDSS Stripe 82 region \citep{ngeow2017} and from 0.013 in $r-z$ to 0.057 in $g-i$ in the globular cluster M5 \citep{vivas2017}. We note further that the slopes of the PC$_{\mathrm{max}}$ relations in the three galaxies vary according to the metallicity. The Bulge has the smallest slope, followed by the LMC and then the SMC. The error on the slopes suggest that the difference in the PC$_{\mathrm{max}}$ slopes between the three galaxies is significant. The color at maximum light is essentially a proxy for amplitude since in these stars temperature fluctuations are more important than radius fluctuations in determining amplitude variations. It is certainly the case in Table~\ref{tab:PCAC} and these results can be used to place strong constraints on stellar evolution/pulsation models of RR Lyrae stars. 

\begin{table*}
\caption{The slopes and intercepts for PC and AC relations for RR Lyrae stars in the Bulge, LMC and SMC (data from OGLE-IV) at minimum, mean and maximum light.}
\centering
\begin{tabular}{c c c c c c c c c c c}
\hline\hline
& Phase & slope(RRab) & intercept(RRab) & $\sigma$(RRab) & slope(RRc) & intercept(RRc) & $\sigma$(RRc)\\
\hline\hline
\multicolumn{8}{c}{Bulge (Using extinction law from \citet{cardelli1989})}\\
\hline
PC&max&	1.359$\pm$0.028	&	0.542$\pm$0.008	&	0.194	&	0.402$\pm$0.04	&	0.388$\pm$0.021	&	0.179 \\
-&min&	0.089$\pm$0.027	&	0.603$\pm$0.007	&	0.185	&	0.483$\pm$0.039	&	0.649$\pm$0.02	&	0.177 \\
AC&max&	-0.378$\pm$0.005	&	0.53$\pm$0.005	&	0.172	&	-0.428$\pm$0.014	&	0.369$\pm$0.007	&	0.174\\
-&min&	0.099$\pm$0.005	&	0.491$\pm$0.005	&	0.183	&	0.546$\pm$0.014	&	0.159$\pm$0.007	&	0.178\\
\hline
\multicolumn{8}{c}{Bulge (Using extinction law from \citet{nataf2013})}\\
\hline
PC&max& 1.397$\pm$0.028 & 0.621$\pm$0.008 & 0.187 & 0.392$\pm$0.037 & 0.449$\pm$0.019 & 0.155\\
-&min& 0.13$\pm$0.028 & 0.669$\pm$0.007 & 0.179 & 0.508$\pm$0.04 & 0.713$\pm$0.021 & 0.168\\
AC&max& -0.373$\pm$0.005 & 0.585$\pm$0.005 & 0.155 & -0.037$\pm$0.016 & 0.267$\pm$0.007 & 0.166\\
-&min& 0.083$\pm$0.006 & 0.563$\pm$0.005 & 0.179 & 0.815$\pm$0.016 & 0.107$\pm$0.008 & 0.178\\
\hline
\multicolumn{8}{c}{LMC}\\
\hline
PC&max&	1.651$\pm$0.017	&	0.681$\pm$0.004	&	0.112	&	0.618$\pm$0.019	&	0.576$\pm$0.01	&	0.076\\
-&min&	0.17$\pm$0.014	&	0.667$\pm$0.004	&	0.094	&	0.786$\pm$0.022	&	0.855$\pm$0.011	&	0.087\\
AC&max&	-0.386$\pm$0.002	&	0.611$\pm$0.002	&	0.088	&	-0.053$\pm$0.011	&	0.295$\pm$0.005	&	0.087\\
-&min&	0.02$\pm$0.003	&	0.609$\pm$0.002	&	0.096	&	0.431$\pm$0.012	&	0.278$\pm$0.005	&	0.092\\
\hline
\multicolumn{8}{c}{SMC}\\
\hline
PC&max&	1.956$\pm$0.039	&	0.755$\pm$0.009	&	0.092	&	0.739$\pm$0.053	&	0.633$\pm$0.025	&	0.054\\
-&min&	-0.128$\pm$0.033	&	0.653$\pm$0.007	&	0.077	&	0.51$\pm$0.068	&	0.776$\pm$0.032	&	0.07\\
AC&max&	-0.375$\pm$0.004	&	0.623$\pm$0.003	&	0.062	&	-0.259$\pm$0.036	&	0.423$\pm$0.019	&	0.065\\
-&min&	0.093$\pm$0.005	&	0.607$\pm$0.004	&	0.076	&	0.389$\pm$0.038	&	0.337$\pm$0.02	&	0.07\\
\hline
\end{tabular}
\label{tab:PCAC}
\end{table*}

\subsection{Theoretical colors of RR Lyrae stars at maximum and minimum light}

\begin{figure*}
\centering
\includegraphics[width=1.0\textwidth,scale=1,trim = 0 0 0 0, clip]{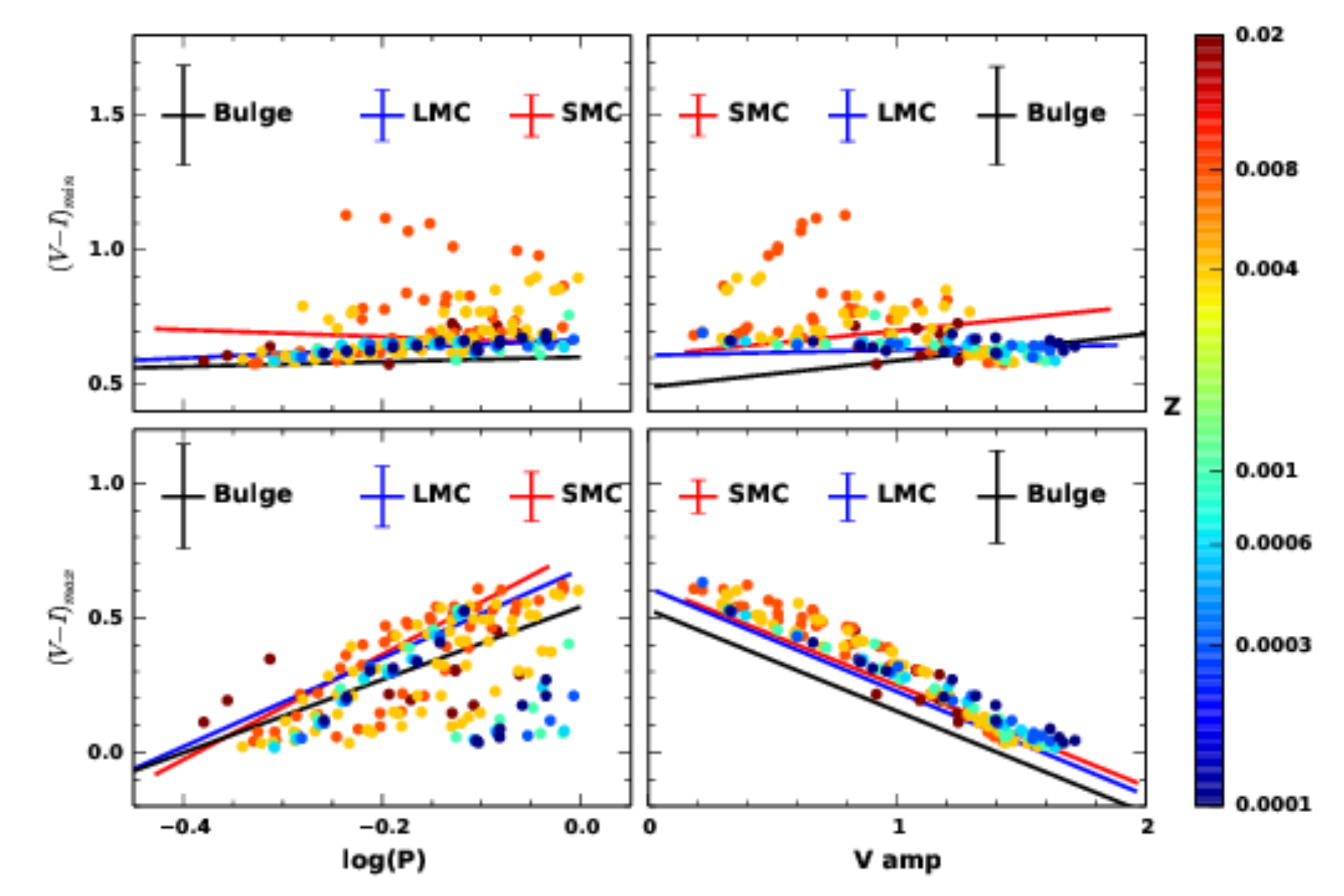}
\caption{PC and AC relations for the RRab models at maximum and minimum light. The period range has been restricted to $\log(P)<0$ for comparison with the observed RRab stars. The lines represent the best fit relations to the PC and AC relations from Bulge, LMC and SMC RRab stars. The standard deviations of the PC and AC plots from the observed data are also plotted on top of each sub-plot.}
\label{fig:PCAC_obsthr}
\end{figure*}

Fig.~\ref{fig:PCAC_obsthr} presents the PC and AC relations at maximum and minimum light obtained from theoretical models with different metal-abundances. The solid lines display these relations in the Bulge, LMC and SMC respectively along with a representative $1\sigma$ error bar. The results of the best-fit linear regressions to the theoretical PC and AC relations are listed in Table~\ref{tab:PCAC_models}. We find the slope of the PC relation at minimum light to be flat for Z=0.02, Z=0.001, Z=0.0006 and Z=0.0003, within 3$\sigma$ uncertainties. We note that the typical mean metallicity of RR Lyrae stars in the LMC is $[Fe/H]=-1.48\pm0.03$ dex on the Harris metallicity scale \citep{clementini2003}, equivalent to a mean metallicity of -1.25$\pm$0.07 \citep{smolec2005} on the High Dispersion Spectroscopy (HDS) scale (or Z=0.0006 with the source for standard abundances from \citet{asplund2005}\footnote{\url{http://astro.wsu.edu/models/calc/XYZ.html} \label{footnote 1}}) . Similarly, the average metal-abundance of RR Lyrae stars in the SMC is much smaller, Z=0.0003. The theoretical models predict a flat PC relation for the RRab stars in the LMC and SMC at minimum light. Interestingly, the PC relation at minimum light is slightly negative for both observed RR Lyrae stars in the SMC and the model Z=0.0003 $-$ this is observed neither in Bulge/LMC nor with the other model compositions.

\begin{table*}
\caption{The slopes and intercepts for PC and AC relations for the RRab models. The period range has been restricted to $\log(P)<0$ for comparison with the observed RRab stars.}
\centering
\begin{tabular}{c c c c c c c c}
\hline\hline
Model&Phase & \multicolumn{3}{c}{PC} & \multicolumn{3}{c}{AC}\\
\hline \hline
& & slope & intercept & $\sigma$ & slope & intercept & $\sigma$ \\
\hline
Z=0.02 & max&	0.351$\pm$0.313	&	0.314$\pm$0.067	&	0.088 & -0.485$\pm$0.147	&	0.775$\pm$0.162	&	0.06\\
& min&	0.351$\pm$0.126	&	0.726$\pm$0.027	&	0.035 &	0.003$\pm$0.126	&	0.657$\pm$0.139	&	0.051\\
Z=0.008 &max&	1.945$\pm$0.191	&	0.694$\pm$0.037	&	0.1 & -0.483$\pm$0.018	&	0.775$\pm$0.017	&	0.042\\
& min&	0.82$\pm$0.266	&	0.894$\pm$0.052	&	0.14 & -0.21$\pm$0.055	&	0.934$\pm$0.053	&	0.133\\
Z=0.004 &max&	1.52$\pm$0.191	&	0.537$\pm$0.034	&	0.122 & -0.453$\pm$0.014	&	0.743$\pm$0.014	&	0.039\\
& min	&0.609$\pm$0.106	&	0.788$\pm$0.019	&	0.068 & -0.126$\pm$0.024	&	0.815$\pm$0.026	&	0.07\\
Z=0.001 &max&	0.234$\pm$0.497	&	0.255$\pm$0.09	&	0.158 & -0.415$\pm$0.024	&	0.691$\pm$0.029	&	0.03\\
&min&	0.17$\pm$0.109	&	0.666$\pm$0.02	&	0.035 & -0.054$\pm$0.026	&	0.701$\pm$0.031	&	0.032\\
Z=0.0006 &max&	0.333$\pm$0.563	&	0.262$\pm$0.108	&	0.151 &	-0.401$\pm$0.013	&	0.684$\pm$0.017	&	0.015\\
&min&	0.137$\pm$0.074	&	0.663$\pm$0.014	&	0.02 &	-0.045$\pm$0.014	&	0.693$\pm$0.017	&	0.015\\
Z=0.0003 &max&	-0.55$\pm$0.263	&	0.138$\pm$0.069	&	0.151 &	-0.397$\pm$0.015	&	0.687$\pm$0.018	&	0.021\\
&min&	-0.037$\pm$0.036	&	0.637$\pm$0.01	&	0.021 &	-0.033$\pm$0.011	&	0.681$\pm$0.014	&	0.016\\
Z=0.0001 &max&	-0.341$\pm$0.635	&	0.176$\pm$0.094	&	0.14 &	-0.349$\pm$0.017	&	0.641$\pm$0.022	&	0.022\\
&min&	0.25$\pm$0.052	&	0.682$\pm$0.008	&	0.011 &	-0.005$\pm$0.016	&	0.656$\pm$0.021	&	0.021\\
\hline
\end{tabular}
\label{tab:PCAC_models}
\end{table*}

\subsection{Period-color-metallicity relation at maximum light}
\label{sec:PCZ}

We note from Table~\ref{tab:PCAC} that the slope of PC relation at maximum light increases as we go from the Bulge to LMC to SMC. Therefore, we investigate metallicity dependence on PC relations at maximum, mean and minimum light. The photometric metallicities ($[Fe/H]$) values are estimated from the light curves of the RRab stars using the relation from \citet{smolec2005}:
\begin{equation}
[Fe/H] = -3.142 -4.902P + 0.824\phi_{31}
\end{equation}
where $P$ is the period in days of the RRab star and $\phi_{31}$ is the Fourier phase parameter listed in Table~\ref{tab:Observed}. This relation provides the metallicity estimates on the HDS scale. We find the median metallicity of LMC to be $-1.267\pm0.002$ dex on the HDS scale. As mentioned earlier, \citet{clementini2003} found a mean metallicity of $-1.48\pm0.03$ dex on the Harris scale which when converted to the HDS scale results in a mean metallicity of $-1.25\pm0.07$ dex \citep{smolec2005}. The median metallicity of SMC is $-1.491\pm0.006$ dex and that of Bulge is $-1.012\pm0.004$ dex on the HDS scale using this relation. Using these median metallicities, we conclude that Z=0.001 for Bulge, Z=0.0006 for LMC and Z=0.0003 for SMC and as such, we use these models for comparison in Figs.~\ref{fig:PCAC_rrab} and \ref{fig:PCAC_rrc}.

Table \ref{tab:PCZ} summarises the $PC-[Fe/H]$ relations for RRab stars in the Bulge, LMC and SMC at minimum, mean and maximum light. The metallicity dependence of the PC relation increases as we go from minimum to maximum light, suggesting that the mean light results are an average of the results at max/min light or of the various pulsation phases. A multiphase approach over multiple wavelengths will be carried out in near-future to investigate this further. To determine the statistical significance of the variables, we check if the p-value of the t-test for the significance of the additional variable ($Pr(\textgreater|t|)$) is less than 0.05. All the variables in the $PCZ$ relations are found to be significant for all the cases except for the $[Fe/H]$ term in the case of LMC at minimum light. 

\begin{table}
\caption{$PC-[Fe/H]$ relations for the OGLE-IV RRab stars in the Bulge, LMC and SMC from OGLE-IV at minimum, mean and maximum light.}
\centering
\scalebox{0.85}{
\begin{tabular}{c c c c c}
\hline\hline
\multicolumn{5}{c}{$(V-I) = a + b\log(P) + c[Fe/H]$}\\
\hline\hline
Phase & $a$ & $b$ & $c$ & $\sigma^*$\\
\hline
\multicolumn{5}{c}{Bulge}\\
\hline
min & 0.571$\pm$0.012 & 0.081$\pm$0.030 & -0.029$\pm$0.007 & 0.181\\
mean & 0.649$\pm$0.011 & 0.622$\pm$0.026 & 0.050$\pm$0.006 & 0.163\\
max & 0.793$\pm$0.012 & 1.608$\pm$0.030 & 0.189$\pm$0.007 & 0.183\\
\hline
\multicolumn{5}{c}{LMC}\\
\hline
min & 0.6688$\pm$0.0074 & 0.1868$\pm$0.0171 & -0.0002$\pm$0.0038 & 0.0909\\
mean & 0.741$\pm$0.006 & 0.646$\pm$0.014 & 0.054$\pm$0.003 & 0.076\\
max & 0.901$\pm$0.008 & 1.732$\pm$0.018 & 0.171$\pm$0.004 & 0.095\\
\hline
\multicolumn{5}{c}{SMC}\\
\hline
min& 0.482$\pm$0.020 & -0.345$\pm$0.048 & -0.081$\pm$0.008 & 0.077\\
mean & 0.712$\pm$0.011 & 0.605$\pm$0.028 & 0.020$\pm$0.005& 0.046\\
max & 0.851$\pm$0.020 & 1.899$\pm$0.048& 0.084$\pm$0.008& 0.078\\
\hline
\end{tabular}}
\begin{tablenotes}
	\small
	\item $^{*}$Standard deviation or internal dispersion.    
\end{tablenotes}
\label{tab:PCZ}
\end{table}

\section{Discussion and Conclusions}
\label{sec:results}

We have carried out a detailed light curve analysis for the largest available dataset of RR Lyrae stars in the Bulge, LMC and SMC from the OGLE-IV survey using the Fourier decomposition technique  and compared the results with the most recent stellar pulsation models of RR Lyrae stars from \citet{marconi2015}. The models show a decrease in amplitude with an increase in wavelength, except for a few period ranges where $U_{amp}<B_{amp}$, depending on the effective temperature of the RR Lyrae star. This is consistent with observations, albeit the mean amplitudes from the models are slightly higher than those from observations $-$ an increase in the mixing length parameter can cause a decrease in the pulsation amplitudes \citep{criscienzo2004b}. Also, the uncertainties on the assumed convective efficiency affect the pulsation amplitudes of the theoretical light curves \citep{fiorentino2007}. An investigation of the variation of Fourier parameters with mass predicts a decrease in Fourier amplitude parameters and an increase in the Fourier phase parameters with an increase in mass for a given period range, especially in the $K$-band. The availability of the NIR RR Lyrae data in the near future would be useful for providing constraints on the M-L combinations of the RR Lyrae stars. The variation of Fourier parameters with wavelength presents a decrease in amplitude parameters and an increase in phase parameters with an increase in wavelength, for a given period. The scatter in the amplitude parameters decreases as we go from optical to infrared, given that the metallicity effects are less at longer wavelengths.

The observed Fourier parameters of RR Lyrae stars are in reasonable agreement with those obtained from the models $-$ with a better consistency in the infrared bands. For the long-period range $0 < \log(P) < 0.2$, models show marginal inconsistency in phase parameters at optical wavelengths. We found a subset of 25 RRab stars from the LMC with I-band light curves that match well with models. These subset of models were used to obtain an average distance modulus to LMC of $18.51\pm0.07$ mag, which is in good agreement with published results.

We study the period-color and amplitude-color relations at minimum and maximum light to understand the interaction of the stellar photosphere with the hydrogen ionisation front. While the PC$_{\mathrm{min}}$ slope is nearly-flat for Bulge RRab stars and consistent with previous results, it has a small but significant positive/negative slope for LMC/SMC RRab stars. However, the change in slope from minimum to maximum light is significant and thereby, the theory of the interaction of stellar photosphere and hydrogen ionisation front is consistent. Unlike their fundamental mode counterparts, the RRc stars show a smaller difference in PC slope from min/max $-$ this is consistent with the theory of the HIF$-$stellar photosphere interaction because these stars are hotter and so the HIF and stellar photosphere engagement occurs at a much higher temperature in a range where Saha ionization equilibrium is more sensitive to temperature.

Using the photometric median metallicities, we find Z=0.001 for Bulge, Z=0.0006 for LMC and Z=0.0003 for SMC and use these models for comparison with the observations. The models predict a flat PC$_{\mathrm{min}}$ at minimum light for Z=0.02, Z=0.001, Z=0.0006 and Z=0.0003. It is interesting to note that Z=0.0003 predicts a slightly negative slope for PC$_{\mathrm{min}}$, similar to that observed in SMC. We, therefore, suggest that PC$_{\mathrm{min}}$ may be used as a constraint for models. The metallicity dependence of the PC relations increases as we go from minimum to maximum light. At maximum light, the PC relation slope increases from Bulge to LMC to SMC. The results of PC and AC relations in both theory and observations are found to be consistent with the previous works and the theory of the interaction of stellar photosphere and hydrogen ionization front. The multi-wavelength light curve analysis of fundamental-mode RR Lyrae stars has been carried out extensively for the first-time in the present analysis using both theoretical models and observed light curves. Although our results suggest an overall consistency of the models with the observations, there are cases of discrepancies such as the higher amplitudes at optical bands and the sensitivity of convection towards the redder edge of the instability strip that need further investigation. A smoother grid of models along with variation in the mixing length, viscosity etc. should result in a better agreement between the models and observations. The results of this work can provide stringent constraints for the theoretical stellar pulsation codes that incorporate static atmosphere models to generate RR Lyrae light curves at multiple wavelengths. 

\section*{Acknowledgements}
The authors thank the referee for useful comments and suggestions that improved the quality of the manuscript. SD acknowledges the INSPIRE Junior Research Fellowship vide Sanction Order No. DST/INSPIRE Fellowship/2016/IF160068 under the INSPIRE Program from the Department of Science \& Technology, Government of India. HPS and SMK thank the Indo-US Science and Technology Forum for funding the Indo-US virtual joint networked centre on ``Theoretical analyses of variable star light curves in the era of large surveys''. AB acknowledges the research grant \#11850410434, awarded by the National Natural Science Foundation of China through a Research Fund for International Young Scientists. AB, SMK and MM acknowledge support by the Munich Institute for Astro- and Particle Physics (MIAPP) of the DFG cluster of excellence ``Origin and Structure of the universe'' during their stay at the MIAPP workshop on ``Extragalactic Distance Scale in the Gaia Era''. The authors acknowledge the contribution of Brett Meerdink and Matthew Sodano of SUNY Oswego in testing the robustness of the negative slope of PC$_\textrm{min}$ from RRab stars in SMC.

\bibliographystyle{mnras}

\appendix
\section{Fourier interrelations}
\begin{table*}
\caption{The Fourier interrelations in multiple bands for the models Z=0.001, Z=0.0006 and Z=0.0003.}
\centering
\scalebox{0.8}{
\begin{tabular}{c c c c c c c c c c c c c c}
\hline\hline
$\lambda_1-\lambda_2$ & FP &  \multicolumn{4}{c}{Z = 0.001} & \multicolumn{4}{c}{Z = 0.0006} & \multicolumn{4}{c}{Z = 0.0003}\\
& & $\beta^a$ & $\alpha^b$ & $\sigma^c$ & $r^d$ & $\beta^a$ & $\alpha^b$ & $\sigma^c$ & $r^d$ & $\beta^a$ & $\alpha^b$ & $\sigma^c$ & $r^d$\\
\hline\hline
V-I & $R_{21}$ & 0.806$\pm$0.058 & 0.063$\pm$0.032 & 0.045 & 0.948 & 0.755$\pm$0.051 & 0.07$\pm$0.028 & 0.053 & 0.952 & 0.795$\pm$0.047 & 0.072$\pm$0.029 & 0.042 & 0.965\\
& $R_{31}$ & 0.884$\pm$0.04 & 0.017$\pm$0.013 & 0.024 & 0.978 & 0.938$\pm$0.038 & -0.002$\pm$0.012 & 0.026 & 0.982 & 1.013$\pm$0.046 & -0.02$\pm$0.015 & 0.022 & 0.979\\
& $\phi_{21}$ & 1.031$\pm$0.05 & 0.372$\pm$0.169 & 0.227 & 0.976 & 1.0$\pm$0.043 & 0.42$\pm$0.144 & 0.197 & 0.982 & 1.197$\pm$0.062 & -0.198$\pm$0.215 & 0.197 & 0.974\\
& $\phi_{31}$ & 0.926$\pm$0.046 & 1.12$\pm$0.173 & 0.347 & 0.982 & 0.939$\pm$0.058 & 1.038$\pm$0.208 & 0.399 & 0.973 & 0.938$\pm$0.043 & 1.01$\pm$0.152 & 0.374 & 0.982\\
I-J & $R_{21}$ & 0.64$\pm$0.11 & 0.097$\pm$0.056 & 0.073 & 0.779 & 0.672$\pm$0.091 & 0.066$\pm$0.044 & 0.076 & 0.839 & 0.495$\pm$0.091 & 0.168$\pm$0.05 & 0.067 & 0.763\\
& $R_{31}$ & 0.76$\pm$0.076 & 0.023$\pm$0.023 & 0.042 & 0.905 & 0.831$\pm$0.065 & 0.001$\pm$0.02 & 0.043 & 0.936 & 1.03$\pm$0.061 & -0.048$\pm$0.018 & 0.03 & 0.965\\
& $\phi_{21}$ & 0.96$\pm$0.052 & 0.612$\pm$0.183 & 0.247 & 0.972 & 0.944$\pm$0.033 & 0.611$\pm$0.111 & 0.205 & 0.987 & 0.937$\pm$0.06 & 0.689$\pm$0.222 & 0.247 & 0.963\\
& $\phi_{31}$ & 1.121$\pm$0.072 & 0.668$\pm$0.163 & 0.383 & 0.975 & 1.042$\pm$0.076 & 0.681$\pm$0.207 & 0.55 & 0.958 & 1.105$\pm$0.099 & 0.768$\pm$0.246 & 0.474 & 0.952\\
J-K & $R_{21}$ & 0.883$\pm$0.146 & -0.086$\pm$0.063 & 0.08 & 0.79 & 0.751$\pm$0.1 & -0.026$\pm$0.04 & 0.067 & 0.843 & 0.64$\pm$0.194 & -0.001$\pm$0.086 & 0.092 & 0.585\\
& $R_{31}$ & 0.642$\pm$0.123 & -0.002$\pm$0.032 & 0.056 & 0.745 & 0.653$\pm$0.091 & -0.004$\pm$0.023 & 0.053 & 0.832 & 0.9$\pm$0.086 & -0.065$\pm$0.023 & 0.045 & 0.915\\
& $\phi_{21}$ & 0.916$\pm$0.054 & 0.788$\pm$0.205 & 0.282 & 0.967 & 0.862$\pm$0.066 & 0.951$\pm$0.256 & 0.342 & 0.944 & 0.933$\pm$0.049 & 0.679$\pm$0.194 & 0.304 & 0.973\\
& $\phi_{31}$ & 0.844$\pm$0.052 & 1.318$\pm$0.107 & 0.329 & 0.964 & 0.835$\pm$0.057 & 1.315$\pm$0.135 & 0.392 & 0.957 & 0.875$\pm$0.059 & 1.318$\pm$0.157 & 0.442 & 0.958\\

\hline
\end{tabular}}
\begin{tablenotes}
	\small
	\item $^{a}$Slope with associated error.
	\item $^{b}$Intercept with associated error.  
	\item $^{c}$Standard deviation or internal dispersion.
	\item $^{d}$Pearson's correlation coefficient.          
\end{tablenotes}
\label{tab:fourier_interrelation}
\end{table*}

\begin{figure}
\centering
\includegraphics[scale=1,trim = 0 0 0 0, clip]{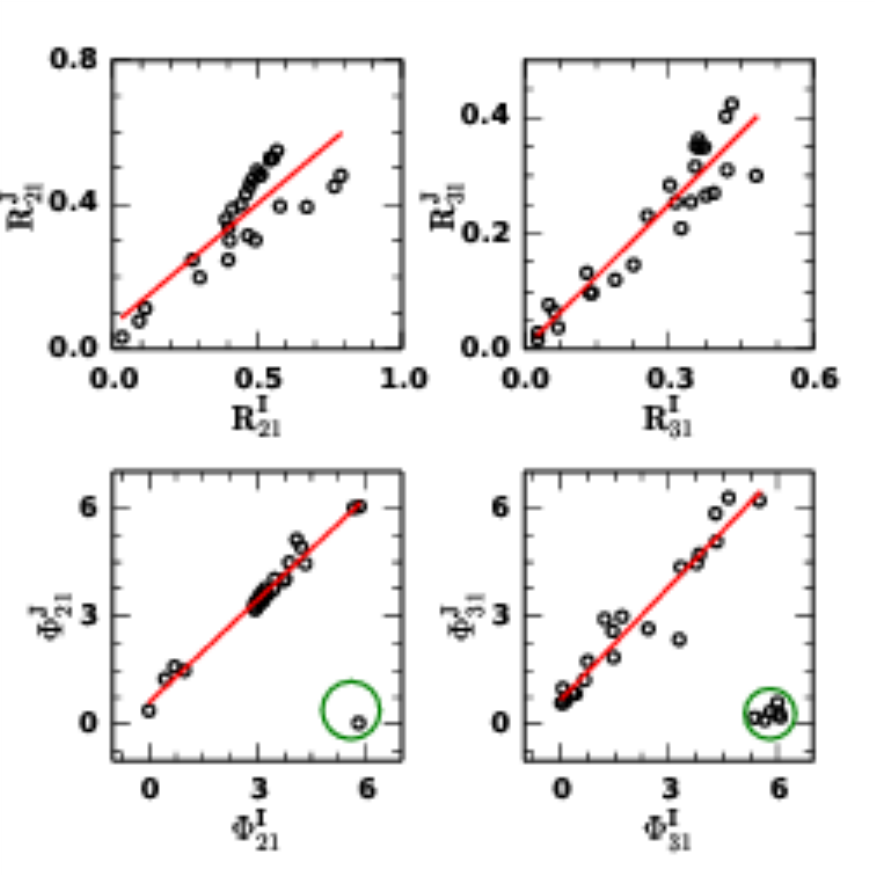}
\caption{The Fourier interrelations between I and J with chemical composition Z=0.0006.}
\label{fig:interrelation_Z0.0006}
\end{figure}

There is a dearth of light curve data for RR Lyrae stars in the NIR bands $-$ usually we have single-epoch observations. Even in case of multi-epoch observations in the NIR bands, the number of observations are typically not enough to calculate precise Fourier parameters. From the comparison of the theoretical and observed Fourier parameters (Sec.~\ref{sec:fourier_obsthr}), we find that albeit a few period ranges where there are slight discrepancies, the observed Fourier parameters match quite well with those from the models. This helps us in predicting the Fourier parameters in the other bands after deriving the Fourier interrelations using the following transformation equations:
\begin{equation}
\begin{aligned}
R_{21}^{\lambda_2} &= \alpha + \beta{R_{21}^{\lambda_1}}, R_{31}^{\lambda_2} &= \alpha + \beta{R_{31}^{\lambda_1}}, \\
\phi_{21}^{\lambda_2} &= \alpha + \beta{\phi_{21}^{\lambda_1}}, \phi_{31}^{\lambda_2} &= \alpha + \beta{\phi_{31}^{\lambda_1}},
\label{eq:interrelation}
\end{aligned}
\end{equation}
where $\lambda_1 < \lambda_2$. Fig.\ref{fig:interrelation_Z0.0006} shows a representative plot of the Fourier interrelations for Z=0.0006 from I-J bands. The green circle encloses a set of ``outliers'' which have been removed prior to obtaining the transformation equations. The set of equations have been summarised in Table~\ref{tab:fourier_interrelation} for Z=0.001, Z=0.0006 and Z=0.0003. The internal dispersion decreases as we go from $R_{21}$ to $R_{31}$ and increases as we go from $\phi_{21}$ to $\phi_{31}$ and it increases as we go from optical to near-infrared bands, for all the cases. This paper provides the transformation equations for Fourier parameters in (V-I), (I-J) and (J-K) bands but it can easily be obtained for the rest of the bands.

\label{lastpage}

\end{document}